\documentclass[10pt,aps,prb,twocolumn,amsmath,amssymb,superscriptaddress]{revtex4-1}
\usepackage{amsmath}
\usepackage{graphicx}
\usepackage{color}
\usepackage{multirow}
\usepackage[caption = false]{subfig}
\usepackage{array}
\usepackage{threeparttable}
\usepackage{mathtools}
\usepackage{siunitx}
\usepackage{centernot}
\usepackage{tikz}
\tikzset{font={\fontsize{11pt}{12}\selectfont}}
\usetikzlibrary{shapes,arrows}
\usepackage{lipsum}
\makeatletter

\renewcommand*{\p@subsection}{}

\renewcommand*{\p@subsubsection}{}
\makeatother

\usepackage{hyperref}
\hypersetup{
	linkcolor=blue,          
	citecolor=blue,        
	urlcolor=blue,           
}

\newcommand{\ket}[1]{|#1\rangle}

\newcommand{\inner}[2]{\langle#1|#2\rangle}

\newcommand{\expecth}[3]{\langle#1|#2|#3\rangle}

\begin{document}
\title{Structure and dynamics of electron-phonon coupled systems using neural quantum states}

\author{Ankit Mahajan}
\affiliation{Department of Chemistry, Columbia University, New York, NY 10027, United States}

\author{Paul J. Robinson}
\affiliation{Department of Chemistry, Columbia University, New York, NY 10027, United States}
\affiliation{Department of Chemistry and Chemical Biology, Harvard University, Cambridge, Massachusetts 02138, United States}

\author{Joonho Lee}
\affiliation{Department of Chemistry, Columbia University, New York, NY 10027, United States}
\affiliation{Department of Chemistry and Chemical Biology, Harvard University, Cambridge, Massachusetts 02138, United States}

\author{David R. Reichman}
\email{drr2103@columbia.edu}
\affiliation{Department of Chemistry, Columbia University, New York, NY 10027, United States}

\begin{abstract}
In this work, we use neural quantum states (NQS) to describe the high-dimensional wave functions of electron-phonon coupled systems. We demonstrate that NQS can accurately and systematically learn the underlying physics of such problems through a variational Monte Carlo optimization of the energy with minimal incorporation of physical information even in highly challenging cases. We assess the ability of our approach across various lattice model examples featuring different types of couplings. The flexibility of our NQS formulation is demonstrated via application to \textit{ab initio} models parametrized by density functional perturbation theory consisting of electron or hole bands coupled linearly to dispersive phonons. We compute accurate real-frequency spectral properties of electron-phonon systems via a novel formalism based on NQS. Our work establishes a general framework for exploring diverse ground state and dynamical phenomena arising in electron-phonon systems, including the non-perturbative interplay of correlated electronic and electron-phonon effects in systems ranging from simple lattice models to realistic models of materials parametrized by \textit{ab initio} calculations.   
\end{abstract}
\maketitle

\section{Introduction}
The interaction between bosonic fields with charged and neutral carriers can lead to the formation of emergent quasiparticles with greatly altered properties.  For example, transport measurements carried out on lightly doped or photoexcited carriers in inorganic and organic semiconductors have been linked to the formation of polarons,\cite{coropceanu2007charge,zhu2015charge,schilcher2021,tulyag2023room} quasiparticles composed of the carrier enveloped by a cloud of phonons.  Features such as kinks and satellites in photoemission spectra at well-defined phonon frequencies offer additional direct evidence of polaronic physics.\cite{li2020strong,verdi2017origin,moser2013tunable,kang2018holstein,setvin2014direct} Moreover, electron-phonon (eph) interactions can dramatically influence the interactions of electrons with other electrons. Perhaps the most important example of this is superconductivity in simple metals, embodied in the Bardeen-Cooper-Schrieffer (BCS) theory, where Cooper pairs formed by phonon-mediated attraction between electrons lead to an instability of the Fermi sea and the formation of an emergent superconducting state.\cite{bardeen1957theory}  The role of strong eph interactions in unconventional superconductivity remains a subject of debate.\cite{lanzara2001evidence,alexandrov2002frohlich,giustino2008small,yin2013correlation,karakuzu2017superconductivity,costa2018phonon,luo2022electronic,ly2023comparative,zhang2023bipolaronic}

Since Landau proposed the concept of the self-trapping of an electron via lattice distortions, there has been extensive theoretical investigation into eph interactions.\cite{alexandrov2010advances,franchini2021polarons,giustino2017electron} This exploration has entailed the development of simple lattice and continuum models, such as those pioneered by Holstein\cite{holstein1959studies,holstein1959studies2} and Fr\"ohlich,\cite{frohlich1954electrons} alongside methods to solve the resulting eph problem. Various strategies for tackling these models include path integral\cite{feynman1955slow,kornilovitch1998continuous,beyl2018revisiting,lee2021constrained,cohen2022fast,malkaruge2023comparative,li2023suppressed} and diagrammatic methods,\cite{prokof1998polaron,berciu2006green} variational\cite{lee1953motion,toyozawa1961self,brown1997variational,alder1997variational,ohgoe2014variational,hohenadler2004quantum,karakuzu2017superconductivity,ferrari2020variational} and perturbative techniques,\cite{marsiglio1995pairing,hohenadler2003spectral} density matrix renormalization group (DMRG) approaches\cite{jeckelmann1998density}, as well as various flavors of exact diagonalization\cite{bonvca1999holstein,wang2020zero}. Although some of these approaches are numerically exact in some domains, their applicability across different types of Hamiltonians and parameter ranges is constrained in practice due to potentially high computational costs. Recently, significant progress has been made in constructing and studying \textit{ab initio} models based on density functional perturbation theory (DFPT).\cite{baroni2001phonons,sjakste2015wannier,giustino2017electron,ponce2016epw,zhou2021perturbo,verdi2015frohlich} Methods initially developed for the investigation of lattice models are now being adapted to this \textit{ab initio} setting.\cite{lee2021facile,lafuente2022unified} Systematically improvable calculations towards the exact limit have not yet been reported. In addition to allowing quantitative comparison with experiments and predictive power, such calculations also help assess the accuracy of approximate but more affordable computational methods.

\begin{figure*}
   \centering
   \includegraphics[width=0.9\textwidth]{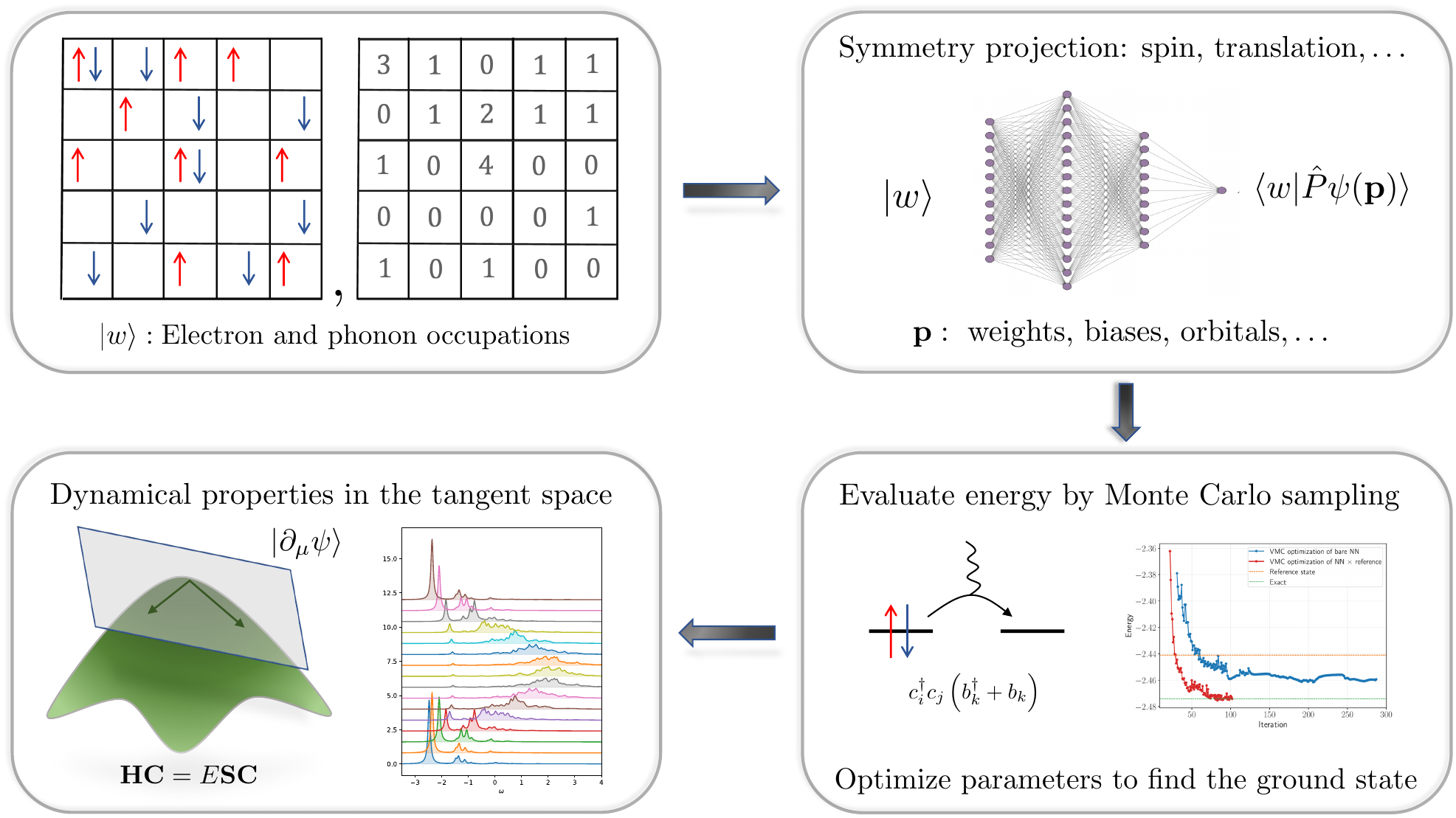}
   \caption{Schematic illustration of the neural quantum state (NQS) approach for electron-phonon coupled systems showing steps in the VMC sampling, optimization, and calculations of dynamical properties.}
   \label{fig:schematic}
\end{figure*}

Neural quantum states (NQS) present a promising avenue for addressing these challenges. Due to their universal approximation properties, neural networks (NN) have emerged as versatile wave function ansatzes for many-body systems.\cite{carleo2017solving} Both empirical and theoretical\cite{gao2017efficient,sharir2020deep} evidence suggests that NN can efficiently represent physically relevant quantum states in non-trivial interacting systems. Leveraging technological developments in machine learning and via the use of Monte Carlo sampling, NQS can often be evaluated and optimized efficiently, allowing one to systematically converge answers close to the exact limit by increasing the number of parameters. Examples of recent successes of the NQS approach include the investigation of low-lying states and dynamics of spin and fermion models on both lattices\cite{carleo2017solving,nomura2017restricted,choo2018symmetries,choo2019two,luo2019backflow,vicentini2019variational,szabo2020neural,robledo2022fermionic} and in continuum settings.\cite{hermann2020deep,pfau2020ab} While a few studies have considered the use of NQS for eph coupled systems,\cite{nomura2020machine,rzadkowski2022artificial} their applicability and potential for detailed investigations of such systems remains largely unexplored. Extracting dynamical quantities from NQS is also an area of active research. The spectral properties of electronic and spin systems have been obtained using the correction vector method,\cite{hendry2019machine} Chebyshev expansion,\cite{hendry2021chebyshev}, and real-time evolution.\cite{mendes2023highly} The development of accurate real-frequency spectral properties is especially important in eph coupled systems, where sharp spectral features that are difficult to capture by analytic continuation at particular phonon frequencies often arise. 

The flexibility and accuracy of NQS come at the expense of having to perform stochastic nonlinear optimization, presenting challenges more severe than those encountered in deterministic methods.\cite{cai2018approximating,westerhout2020generalization,szabo2020neural} Good results require careful development of heuristics for the choice of representations and learning strategies, which have to be tailored to the problem domain. In this work, we present our efforts to address these challenges for eph systems.  We find that the NQS approaches we put forward here are competitive with the best exact approaches for standard polaron problems, while providing facile access to highly accurate spectral properties and the ability to flexibly treat complex problems such as those with correlated electrons and {\em ab initio} parametrized Hamiltonians.
In the following sections, we present the details of Hamiltonians, wave functions, and the methods used to calculate their properties (Sec. \ref{sec:theory}). We then present the results of our calculations on a variety of eph systems to demonstrate the utility of our approach (Sec. \ref{sec:results}). Finally, we conclude with a discussion of our results and future directions (Sec. \ref{sec:conclusion}).

\section{Theory}\label{sec:theory}
\subsection{Hamiltonians}\label{sec:hamiltonians}
We consider the general linear coupling eph Hamiltonian which, when written in the momentum basis, is given by

\begin{equation}
   \begin{split}
      H &= \sum_{n,k}\epsilon_{nk} c^{\dagger}_{n,k}c_{n,k} + \sum_{\nu,q}\omega_{\nu q} b^{\dagger}_{\nu,q}b_{\nu,q}\\
      &+ \sum_{\substack{k_1,k_2,k_3\\ijmn}} V_{k_1k_2k_3}^{ijmn} c_{i,k_1}^{\dagger}c_{j,k_2}^{\dagger}c_{m,k_3}c_{n,k_1 + k_2 - k_3} \\
      & + \sum_{mn\nu kq}g_{mn\nu kq} c^{\dagger}_{n,k+q}c_{n,k}(b^{\dagger}_{\nu,-q} + b_{\nu,q}).\\   
   \end{split}
\end{equation}
Here \(c_{n,k}\) are electronic annihilation operators for an electron with crystal momentum \(k\) in band \(n\), and \(b_{\nu q}\) are phonon annihilation operators with crystal momentum \(q\) in band \(\nu\). \(\epsilon_{k}\) are electronic energies, \(V_{k_1k_2k_3}^{ijmn}\) are electron-electron interactions, \(\omega_{\nu, q}\) are phonon energies, and \(g_{mn\nu kq}\) are eph couplings. We have omitted spin indices for brevity. In an \textit{ab initio} setting, we obtain the band energies and interaction terms from DFPT calculations.\cite{giustino2017electron,ponce2016epw,zhou2021perturbo}

\begin{table*}\label{tab:models}
   \caption{Summary of the models on \(n\)-dimensional integer lattices studied in this work. In the Bond and SSH models, \(i\) and \(j\) represent neighboring sites, with \(\langle ij\rangle\) denoting the bond between them, and \(\left\{ij\right\}\) the direction of the bond. \(\mu\) denotes the phonon band index. \(N\) is the number of lattice sites. Sum over repeated indices is implied.}
   \centering
      \begin{tabular}{ccm{2cm}cm{6cm}cm{5cm}cm{2cm}c}
         \hline
         Model &~& \centering Phonon &~& \multicolumn{3}{c}{Electron phonon coupling term \(H_{\text{eph}}\)} &~& \centering Dimensionless &\\
         \cline{5-7}
         &~& \centering degrees of freedom &~& \centering Site basis &~& \centering Momentum basis &~& \centering coupling constant & \\
         \hline
         Holstein &~& \centering 1-d phonons on sites
         &~& \begin{minipage}[b]{\linewidth}
            \centering \begin{equation*}
            gc^{\dagger}_ic_i(b^{\dagger}_i+b_i)
         \end{equation*}\end{minipage} &~& \begin{minipage}[b]{\linewidth}\centering \begin{equation*}
            \dfrac{g}{\sqrt{N}}c^{\dagger}_{k+q}c_k(b^{\dagger}_{-q} + b_q)
         \end{equation*}\end{minipage} &~& \begin{minipage}[b]{\linewidth}\centering\begin{equation*}
            \lambda = \dfrac{g^2}{2\omega t}
         \end{equation*}\end{minipage} &\\
         SSH &~& \centering \(n\)-d phonons on sites &~& \begin{minipage}[b]{\linewidth}\centering\begin{equation*}
            {\begin{split} &g\left(c^{\dagger}_ic_j + c^{\dagger}_jc_i\right)\\
               &\qquad (b^{\dagger}_{i,\left\{ij\right\}}+b_{i,\left\{ij\right\}} - b^{\dagger}_{j,\left\{ij\right\}}-b_{j,\left\{ij\right\}})\end{split}}
         \end{equation*}\end{minipage} &~&  \begin{minipage}[b]{\linewidth}\begin{equation*}
            {\begin{split} \dfrac{2ig}{\sqrt{N}}&\left(\sin(k_{\mu}+q_{\mu}) - \sin(k_{\mu})\right)\\
            & \qquad \qquad c^{\dagger}_{k+q}c_k(b^{\dagger}_{-q,\mu} + b_{q,\mu}) \end{split}}
         \end{equation*}\end{minipage} &~& \begin{minipage}[b]{\linewidth}\centering \begin{equation*}
            \lambda = \dfrac{g^2}{\omega t}
         \end{equation*}\end{minipage} & \\
         Bond &~& \centering 1-d phonons on bonds &~&  \begin{minipage}[b]{\linewidth}\centering \begin{equation*}
            g\left(c^{\dagger}_ic_j + c^{\dagger}_jc_i\right)(b^{\dagger}_{\langle ij\rangle}+b_{\langle ij\rangle})
         \end{equation*}\end{minipage} &~& \begin{minipage}[b]{\linewidth}\centering\begin{equation*}
            {\begin{split} \dfrac{g}{\sqrt{N}} & \left(e^{i(k_{\mu}+q_{\mu})} + e^{-ik_{\mu}}\right)\\
            & \qquad\qquad c^{\dagger}_{k+q}c_k(b^{\dagger}_{-q,\mu} + b_{q,\mu}) \end{split}}
         \end{equation*}\end{minipage} &~& \begin{minipage}[b]{\linewidth}\centering\begin{equation*}
            \lambda = \dfrac{g^2}{2\omega t}
         \end{equation*}\end{minipage} &\\
         \hline
      \end{tabular}
\end{table*}

Phenomenologically, various special cases of this Hamiltonian have been studied in the literature. For polarons, models like the Holstein\cite{holstein1959studies,holstein1959studies2} and eph coupled SSH\cite{su1979solitons} models include only local interactions often with dispersionless phonons. They have the following general form in the site and momentum bases,
 \begin{equation}
   \begin{split}
      H &= H_e + H_{\nu} + H_{\text{eph}},\\
      H_e &= -\sum_{\langle ij\rangle}t(c^{\dagger}_{i}c_{j} + \text{h.c.}),\\
      & = \sum_{k} -2t \left(\sum_{\mu}\cos(k_{\mu})\right) c^{\dagger}_{k}c_{k},\\
      H_{\nu} &= \omega_0\sum_{\nu,i} b^{\dagger}_{\nu,i}b_{\nu,i} = \omega_0\sum_{\nu,q} b^{\dagger}_{\nu,q}b_{\nu,q},\\
   \end{split}
\end{equation}
where \(t\) is the electronic hopping amplitude between neighboring sites \(i\) and \(j\), and \(\omega_0\) is a fixed frequency for all phonon branches. We set \(t=1\), so that all energies are measured in units of \(t\). The use of periodic boundary conditions allows the Hamiltonian to be expressed in a simple form in the momentum basis. In this work, we consider integer lattices viz. one-dimensional chain, two-dimensional square lattice, and three-dimensional cubic lattice.

Electron-phonon interactions are given in Table \ref{tab:models} for the Holstein, Bond, and SSH models. The Holstein Hamiltonian couples phonons on lattice sites to the local density of electrons, referred to as diagonal coupling. The Holstein model is appropriate for describing the coupling of doped or excited charge carriers to high-frequency optical phonons. The SSH model, also known as the Peierls model, on the other hand, couples carrier hopping to vibrations through an off-diagonal coupling. It results from the modulation of hopping parameters by changes in nuclear positions. A variation of this model, termed the Bond model,\cite{sengupta2003peierls} consists of phonon modes situated on bonds rather than lattice sites, which couple directly to the carrier hopping. These polaron models can be extended to study systems with multiple interacting electrons. On-site Hubbard interactions are commonly used for this purpose, where the interaction is given by
\begin{equation}
   H_{\text{Hubbard}} = U\sum_{i}n_{i\uparrow}n_{i\downarrow}.
\end{equation}

While VMC does not have a sign-problem \textit{per se}, states with non-trivial amplitude sign or phase structures can be harder to optimize.\cite{cai2018approximating,westerhout2020generalization,szabo2020neural} The Holstein and Bond models are stoquastic\cite{bravyi2006complexity} in the site basis, whereas the SSH model is not. To see this note that in a basis specified by the position of the electron \(x_e\) and number of phonons at each lattice site \(\left\{\nu_i\right\}\), these Hamiltonians have negative off-diagonal elements when \(t\) and \(g\) are positive. Thus, according to the Perron-Frobenius theorem, these Hamiltonians have a unique ground state with positive amplitudes in this basis. The SSH Hamiltonian does not have this property and, therefore, does not have a sign-definite ground state in the same basis. Furthermore, while the Holstein model maintains stoquasticity in the momentum basis, the Bond and SSH models do not share this attribute. We note that similar observations have been noted in studies using diagrammatic quantum Monte Carlo (DQMC).\cite{prokof2022phonon} 

\subsection{Wave functions}\label{sec:wave_functions}
We will use the notation \(\textbf{n} = (\left\{e_i\right\}, \left\{\nu_i\right\})\), where \(\left\{e_i\right\}\) and \(\left\{\nu_i\right\}\) are electron and phonon occupation numbers, respectively, to denote the basis vectors. Our NN wave function ansatz is given by

\begin{equation}
   \ket{\psi} = \sum_{\textbf{n}}\frac{\exp\left[f(\textbf{n})\right]}{\sqrt{\prod_i \nu_i!}}\ket{\mathbf{n}},\label{eq:nn_wave_function}
\end{equation}
where the function \(f\) is a sum of two neural nets:
\begin{equation}
   f(\textbf{n}) = r(\textbf{n}) + i\phi(\textbf{n}).
\end{equation}
These NNs operate with real parameters and generate real outputs. The function \(\phi\) serves to impart a phase to the wave function. The use of an exponential form, as identified in prior research, aids in the effective representation of wave function amplitudes that may vary significantly across many orders of magnitude. The factor in the denominator involving phonon numbers is included to improve the stability of wave function optimization. It echoes the functional form of coherent states of the harmonic oscillator. In the case of a single site coupled to a phonon mode, the neural net simply has to represent a linear function of the phonon number. Other ways of encoding phonon degrees of freedom in wave functions include continuum representations\cite{ohgoe2014variational} and binary string representations of phonon occupation numbers.\cite{nomura2020machine,jeckelmann1998density} The latter has the advantage that all inputs are binary but a restriction is placed on the total number of phonons on a site.

In this work, we employ multilayer perceptrons (MLPs), which are fully connected feedforward networks.\cite{cybenko1989approximation,goodfellow2016deep} An MLP with \(n\) layers is defined recursively as
 
\begin{equation}
   \mathbf{x}_{i+1} = \sigma(\mathbf{W}_i.\mathbf{x}_i + \mathbf{b}_i), 
\end{equation}
where \(\mathbf{x}_i\) are outputs of hidden neuron layers with \(\mathbf{x}_0\) the input, \(\mathbf{W}_i\) a weight matrix, \(\mathbf{b}_i\) a bias, and \(\sigma\) is the activation function which acts element-wise on the vector input. In this work, we use rectifier (ReLU) activation functions.\cite{goodfellow2016deep} Other network architectures, like convolutional neural networks (CNN),\cite{choo2018symmetries,roth2023high} restricted Boltzmann machines,\cite{carleo2017solving,nomura2020machine,glasser2018neural}and autoregressive neural nets\cite{humeniuk2023autoregressive} have been used in several studies. CNNs, in particular, have the advantage of being inherently translationally invariant. Translational invariance can be imposed on MLP states too, as we discuss next.

The use of symmetry is essential for increasing the efficiency of the representation. Encoding symmetries biases the NN in a way that obviates the work required to learn them from scratch, allowing one to achieve similar accuracy with a smaller NN with fewer parameters. It also enables the targeting of excited states belonging to different irreducible representations (irrep) of the symmetry group. One way to impose symmetries is to generate images of the input under the action of all elements of the symmetry group and average the NN outputs over them
\begin{equation}
   \psi^S(\textbf{n}) = \frac{1}{N}\sum_{g}c_g\psi(g\textbf{n}),\label{eq:translational_symmetry}
\end{equation} 
where \(g\) denotes a group operation and \(c_g\) is its character in the irrep being targeted. This approach has been used in VMC with traditional wave functions\cite{tahara2008variational,becca2017quantum} as well as in studies of NQS states.\cite{roth2023high,reh2023optimizing} Another possibility is to arbitrarily choose one of the equivalent sets of permutations, e.g. the lexicographically smallest one. For polarons, a natural choice is shifting phonon occupations along with the electron in real space.
\begin{equation}
   \psi^S(\textbf{n}) = \psi(T_0\textbf{n}),
\end{equation}
where \(T_0\) denotes the translation operator that shifts the electron to an arbitrarily chosen origin of the lattice. Although this approach is computationally cheaper, we find it is susceptible to converging to local minima during optimization in our numerical experiments. Therefore we use the averaging method in this work. We note that when working in the momentum basis, imposing translational symmetry becomes trivial, because one can simply restrict the Monte Carlo random walk to configurations with a fixed momentum. For polaron and bipolaron problems, the cost of evaluating the dense Hamiltonian in momentum space is comparable to the cost of translational symmetry projection in the site basis. This is not the case for the treatment of many-electron systems, where the inter-electronic interaction incurs a steeper cost in momentum space. Other Abelian symmetries like some point groups (not used here) can be similarly restored. In bipolaron problems, we make use of spin symmetry in addition to translational symmetry to target singlet states. 

In addition to NN states, we also consider physically motivated approximate polaron wave functions for comparison. The ansatz due to Davydov is given by

\begin{equation}
   \ket{\psi_{\text{Davydov}}} = \left(\sum_i \phi_i c_i^{\dagger}\right) \exp \left(-\sum_{\nu}\xi_{\nu}b_{\nu}^{\dagger} - \xi_{\nu}^*b_{\nu}\right)\ket{0},\label{eq:davydov}
\end{equation}
where \(i\) and \(\nu\) denote electron and phonon indices, respectively, and \(\phi_i\) and \(\xi_{\nu}\) are complex variational parameters. This wave function is a product of a coherent state for the phonons and a linear combination of electron creation operators, designed to represent a small polaron localized in real space. We note that an MLP wave function can mimic this form efficiently with a single hidden neuron. This is accomplished by setting the weights connecting the electronic occupations to the hidden neuron to \(\ln \phi_i\) and those for the phonon occupations to \(\xi_{\nu}\), and using a linear activation function. Imposing translational symmetry on the Davydov wave function entangles the electron with the phonons giving rise to the Toyozawa ansatz, 
\begin{equation}
   \ket{\psi_{\text{Toyozawa}}^{k}} = P_{k} \ket{\psi_{\text{Davydov}}},
\end{equation}
where \(P_{k}\) denotes a projector onto the subspace of fixed momentum \(k\), which acts as in Eq. \ref{eq:translational_symmetry}. Translationally symmetrized NN states can thus be thought of as a generalization of the Toyozawa ansatz. Note that this wave function can be evaluated and optimized deterministically at a polynomial cost, but we use it within VMC here. The low computational cost of its evaluation allows us to assess finite-size effects in cases where calculations with NN states are expensive on large systems. Other generalizations of the Davydov ansatz like the global-local\cite{brown1997variational} and delocalized D\(_1\) ansatzes\cite{sun2013delocalized} yield more accurate results compared to Toyozawa, especially for non-local couplings.\cite{leedd1} It would be interesting to explore the possibility of basing NN states on these wave functions in future work.

The ability of our ansatz to capture electron correlation can be demonstrated by constructing Jastrow states as MLPs. Jastrow factors are known to describe Hubbard-like physics very accurately\cite{tahara2008variational,becca2017quantum} and are given by

\begin{equation}
   J(\mathbf{n}) = \exp\left(\sum_{ij}v_{ij}n_in_j\right),
\end{equation}
where \(n_i\) is the occupation number of electrons at site \(i\), and \(v_{ij}\) are variational parameters. One can obtain the product of a pair of occupation numbers using unit weights and a bias of -1 with a ReLU activation function, 
\begin{equation}
   n_1 n_2 = \text{ReLU}(n_1 + n_2 - 1).
\end{equation}
It might appear that to obtain all pairwise products in the Jastrow factors requires \(O(N^2)\) hidden neurons with \(O(N^3)\) parameters, but it is possible to construct an MLP with \(O(N^2)\) hidden neurons as one would expect given the quadratic scaling number of parameters in \(v_{ij}\). This is easiest to see with a two-hidden layer network. One half of the the first hidden layer consists of \(N\) neurons, one for each input. For a given input site one constructs the following quantity as the output of the corresponding hidden neuron:
\begin{equation}
   \theta_i = \sum_j v_{ij}n_j.
\end{equation} 
For convenience, we also copy over the inputs to another set of \(N\) hidden neurons in the first hidden layer. In the second layer, we construct \(N\) products of the form \(n_i\theta_i\) using the construction above. Finally, these outputs are summed and exponentiated to obtain the Jastrow factor in the output neuron. Note that one can encode electron-phonon Jastrow factors\cite{karakuzu2017superconductivity,ohgoe2014variational} in a similar manner. An efficient encoding of Jastrow factors as Restricted Boltzmann Machines has been reported previously.\cite{clark2018unifying}

\subsection{Variational Monte Carlo (VMC)}\label{sec:vmc}
We optimize and calculate the properties of the above wave functions using Monte Carlo sampling. Although VMC has not been used extensively for calculations of eph systems, some studies have employed this technique\cite{karakuzu2017superconductivity,ohgoe2014variational,ferrari2020variational}. VMC is suitable for computing properties of NN wave functions since it only requires the overlap of the state with a walker configuration. Here, we perform random walks in the space of electron and phonon number configurations. An observable \(O\) can be sampled using
\begin{equation}
   \frac{\expecth{\psi}{O}{\psi}}{\inner{\psi}{\psi}} = \sum_{w}                                                                   \frac{|\inner{\psi}{w}|^2}{\inner{\psi}{\psi}}\frac{\expecth{w}{O}{\psi}}{\inner{w}{\psi}},
\end{equation}
where we have combined carrier and phonon coordinates into \(w\). This requires the evaluation of the following local quantity for each walker configuration.
\begin{equation}
   O_L(w) = \frac{\expecth{w}{O}{\psi}}{\inner{w}{\psi}} = \sum_{w'}\expecth{w}{O}{w'}\frac{\inner{w'}{\psi}}{\inner{w}{\psi}},
\end{equation}
where \(w'\) are configurations generated from \(w\) by application of the observable operator. The cost of local energy evaluation in the site basis for a space-local Hamiltonian, like the Holstein model, is the same as the cost of overlap calculation since the number of excitations generated is a small system-size independent constant. For general long-range Hamiltonians, this is no longer the case and local energy evaluation becomes the bottleneck of the VMC calculation. 

We sample walkers \(w\) from the distribution \(\frac{|\inner{w}{\psi}|^2}{\inner{\psi}{\psi}}\) using continuous time sampling, which is a rejection-free sampling technique\cite{sabzevari2018improved}. Starting from the current walker \(w\), we choose the next walker configuration out of the \(w'\) configurations generated during the local energy evaluation with probability proportional to \(|\frac{\inner{w'}{\psi}}{\inner{w}{\psi}}|\). One of the advantages of the VMC approach is that one is not required to truncate the phonon space, as there are no restrictions on the number of phonons sampled. This is in contrast to deterministic methods like DMRG\cite{jeckelmann1998density}. The gradient of the energy with respect to the variational parameters is similarly sampled. The use of backpropagation allows efficient calculation of energy gradients at the same cost as energy.\cite{goodfellow2016deep} Having sampled energies and gradients, the variational parameters can be optimized using gradient-based optimization methods. We use the AMSGrad method, which is a variation of gradient descent with momentum.\cite{reddi2019convergence}  

\subsection{Real-frequency Green's functions and excited states}\label{sec:greens_functions}
To calculate the dynamical properties of the system, we work in the tangent space spanned by the derivatives of the wave function with respect to the variational parameters.\cite{mcweeny1989method} The state corresponding to the \(\mu\)th parameter is given by
\begin{equation}
   \ket{\psi_{\mu}} = \frac{\partial \ket{\psi_0}}{\partial p_{\nu}},
\end{equation}
with \(\ket{\psi_0}\) being the optimized ground state. The resulting basis is nonorthogonal and has linear dependencies. Thus to obtain the excited states and spectral functions, we solve the generalized eigenvalue problem
\begin{equation}
   \mathbf{H}\mathbf{C} = E\mathbf{S}\mathbf{C},
\end{equation} 
where the Hamiltonian and overlap matrices are given by
\begin{equation}
   \begin{split}
      \mathbf{H}_{\mu\nu} &= \expecth{\psi_{\mu}}{H}{\psi_{\nu}},\\
      \mathbf{S}_{\mu\nu} &= \inner{\psi_{\mu}}{\psi_{\nu}}.
   \end{split}\label{eq:lr_vmc}
\end{equation}
The eigenspectrum of this effective Hamiltonian \(\left\{E_{i}, \ket{i}\right\}\) can be used to calculate spectral functions as
 \begin{equation}
   \begin{split}
       A(\omega) &= -\frac{1}{\pi}\text{Im}\expecth{\psi}{\frac{1}{\omega - H + i\eta}}{\psi}\\
       &= -\frac{1}{\pi}\sum_{i}|\inner{\psi}{i}|^2 \frac{\eta}{(\omega-E_i)^2+\eta^2},\\
   \end{split}
\end{equation}
where \(\eta\) is a Lorentzian broadening parameter. The state \(\ket{\psi}\) is usually a physically intuitive excitation on top of the ground state and depends on the particular spectral function of interest. We refer to this method as linear response VMC (LR-VMC). Despite being couched in a linear response framework, we note that LR-VMC can be formally converged to the exact spectral function by increasing the number of parameters in the ground state wave function, thus expanding the tangent space systematically. We show numerical examples of this convergence in Sec. \ref{sec:dynamical_properties}. The rate of this convergence is usually faster for lower-lying states compared to higher energy excitations. A faster convergence could be achieved by using higher derivative states instead of more parameters, but we do not pursue this possibility here.

While it is possible to sample the amplitude square of the ground state to obtain \(\mathbf{H}\) and \(\mathbf{S}\), this can lead to noisy estimates, as \(\ket{\mu}\) states may have support on configurations with vanishing contributions to \(\ket{\psi_0}\). Moreover, noise in the unbiased estimates of these matrix elements leads to a bias in the estimates of the energy spectrum. We instead employ an alternative approach, termed reweighting in Ref. \citenum{li2010variational}, which uses \(\sum_{\nu}|\inner{n}{\psi_{\mu}}|^2\) as the sampling function instead. The two approaches lead to identical results in the limit of infinite sampling, but we found the reweighting method to perform significantly better for a fixed sampling effort. Details and comparisons of these sampling approaches are provided in Appendix \ref{app:lr_vmc_sampling}. Due to stochastic sampling, the estimated metric \(S\) is not necessarily positive definite and needs to be regularized. We diagonalize the sampled metric and throw away states with eigenvalues below a small threshold. 

Symmetry is again imposed with the use of appropriate walkers used in the VMC sampling. The basis states can therefore be thought of as \(\hat{P}\ket{\psi_{\mu}}\), where \(\hat{P}\) is a projector of the symmetry imposed during the sampling. Excited state spectra of different symmetries can be calculated from the optimized ground states in the corresponding sectors. 

\begin{figure*}[t]
   \centering
   \includegraphics[width=\textwidth]{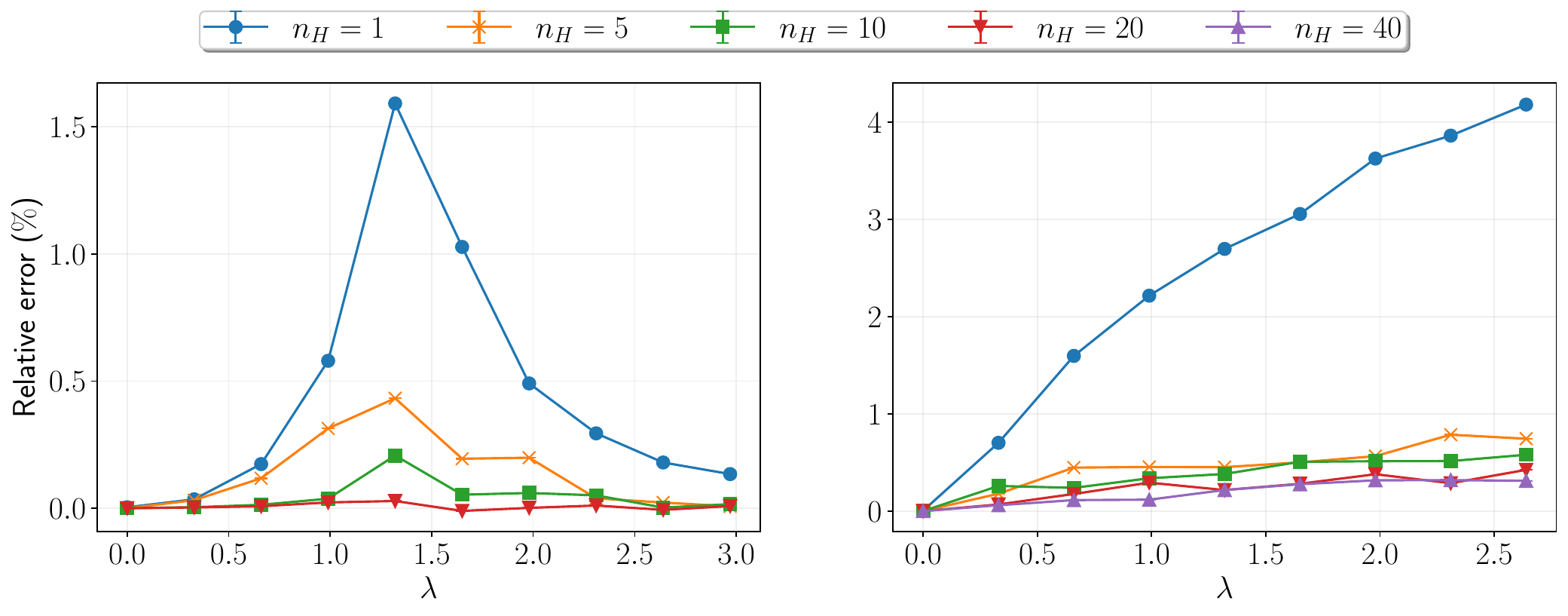}
   \caption{Convergence of error of the NQS ground state energy for the Holstein (left) and Bond (right) model polarons on a 30 site chain with respect to the number of hidden neurons (\(n_H\)) with \(\omega_0=0.5\). We use DMRG energies converged with respect to the number of phonons as a reference.}\label{fig:nh_convergence}
\end{figure*}

This approach of using derivative states has been used with many classic wave function theories, often termed as linear response or equation of motion (EOM) methods.\cite{mcweeny1989method} The Tamm-Dancoff approximation, which is formulated in the tangent space of the Hartree-Fock state, is a well-known example. A linear response DMRG theory has been developed in close analogy to Hartree-Fock.\cite{dorando2009analytic,wouters2013thouless,haegeman2013post} Similarly to NQS, linear response DMRG can also be systematically improved by increasing the bond dimension of the matrix product state. In VMC, an EOM theory based on geminal wave functions was presented in Ref. \citenum{zhao2016equation}. A few studies have obtained dynamical information through the use of a basis constructed from physically relevant excitations on top of the variational ground state.\cite{ido2020charge,ferrari2020variational} Derivative basis states have the advantage of being computationally cheaper for the calculation of the required matrix elements. For example, consider the following element used in the sampling of the Hamiltonian matrix 

\begin{equation}
   \expecth{n}{H}{\psi_{\mu}} = \partial_{\mu}\expecth{n}{H}{\psi_0}.
\end{equation}

One can obtain this matrix element for all \(\mu\) at the same cost scaling as the local energy calculation through reverse mode automatic differentiation. This is in contrast to the bases consisting of excitations on top of the ground state, where in general the cost of computing matrix elements for all excitations scales linearly with the number of excitations. A computational bottleneck of this approach lies in the explicit construction of the \(H\) and \(S\) matrices, which becomes infeasible for a large number of parameters, restricting us to states with fewer than roughly \(10^4\) parameters. This cost can be avoided by using a direct method (not used in this work) outlined in Appendix \ref{app:lr_vmc_sampling}, which only samples the action of these matrices onto vectors. 

\section{Results}\label{sec:results}
We present a numerical analysis of the performance of NQS in polaron, bipolaron, and many-electron systems. First, we consider the ability of these states to represent ground state structure for different kinds of eph coupling. We also calculate the binding energy of the hole polaron in lithium fluoride (LiF) from first principles. In the second part, we assess the accuracy of our LR-VMC approach based on NQS to capture spectral properties of eph systems. We use three-layer MLPs (input, hidden, output) for the radial and phase part of the NQS in all cases, unless stated otherwise. The code used to perform the VMC calculations is available in a public repository.\cite{vmc_code} DMRG results used for reference were obtained using the ITensor library.\cite{Itensor}

\subsection{Ground state properties}\label{sec:ground_state_properties}
\subsubsection{Convergence with number of parameters}\label{sec:polarons}

We start with the question of how many hidden neurons are required to accurately represent the ground state of Holstein and Bond model polarons at different coupling strengths. For the Holstein model, the phase function was set to one, whereas for the Bond model, we use MLPs with identical structures for both the radial and phase parts of the wave function in momentum space. Fig. \ref{fig:nh_convergence} shows the relative percent errors in the ground state energies for a 30 site chain with \(\omega=0.5\). We use DMRG energies converged with respect to the number of phonons as a reference. For the Holstein model, the error is largest for a given number of neurons at intermediate coupling around the self-trapping crossover. This is intuitively sensible since simple weak and strong coupling ansatzes describe the regions away from the crossover point very well. The ansatz with a single hidden neuron, which is equivalent to the Toyozawa wave function, performs very well for the Holstein model. Convergence with respect to the number of hidden neurons is achieved very quickly, with the energies close to exact with just 10 hidden neurons. For the Bond model, on the other hand, the error increases with coupling for a fixed number of hidden neurons. The errors for the same number of hidden neurons are much larger compared to the Holstein model. This model does not exhibit a self-trapping crossover, and the number of hidden neurons required to obtain the same error in ground state energy increases with the size of the coupling. This also coincides with there not being a simple strong coupling ansatz that describes the strong coupling limit of the Bond model. We note that these observations also hold for calculations in the site basis and in higher dimensions. The number of hidden neurons for a given energy accuracy does not scale with the lattice size. 

\begin{figure}[htp]
   \centering
   \includegraphics[width=\columnwidth]{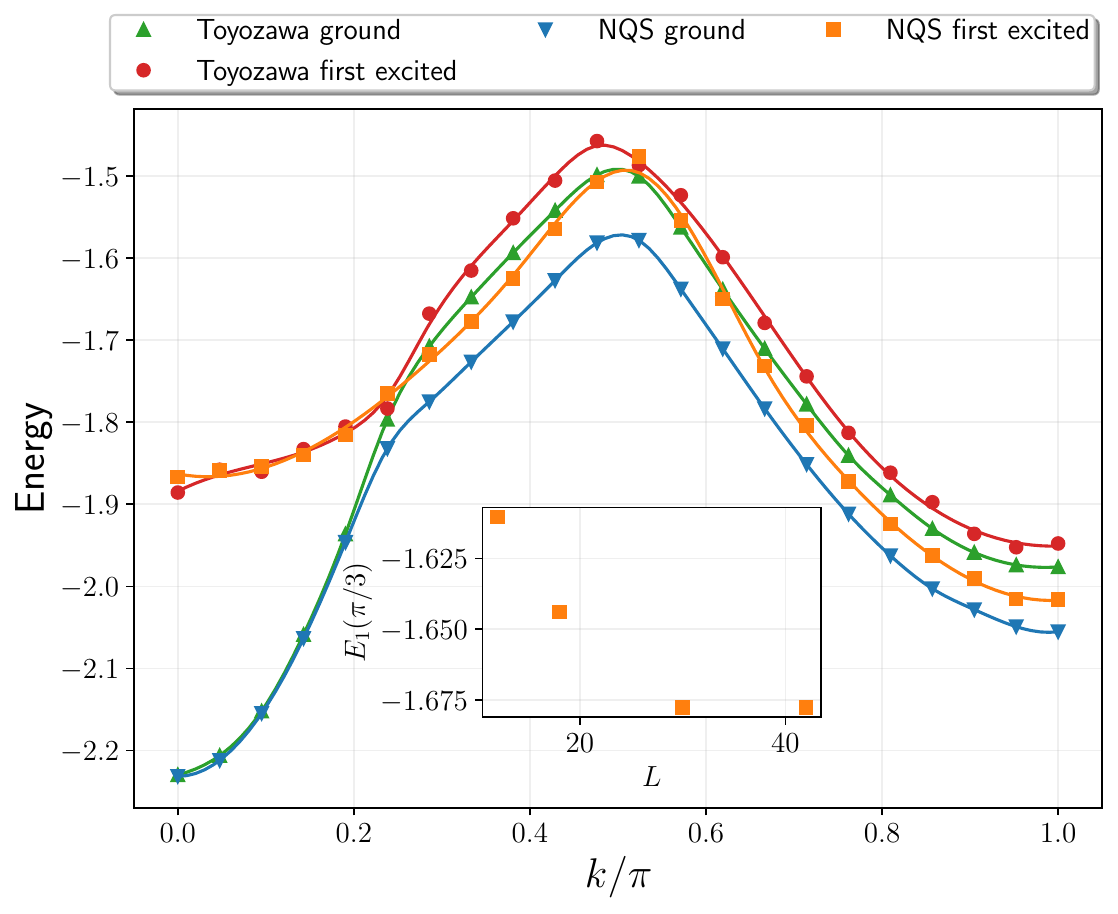}
   \caption{Ground and first excited state bands for a modified Holstein model with dispersive phonons (\(\omega_0=1, t_{\text{ph}}=0.4, \lambda=0.5\)). VMC calculations shown in the main plot were performed on a 42 site chain. Inset shows the convergence of finite size effects in the first excited state energy at \(k=\frac{\pi}{3}\). We also show the bands obtained using the Toyozawa ansatz, which shows substantial deviations from the NQS results, especially close to avoided crossings.}\label{fig:ph_dispersion}   
\end{figure}

\begin{figure*}[t]
   \centering
   \subfloat[Holstein]{
      \includegraphics[width=0.9\columnwidth]{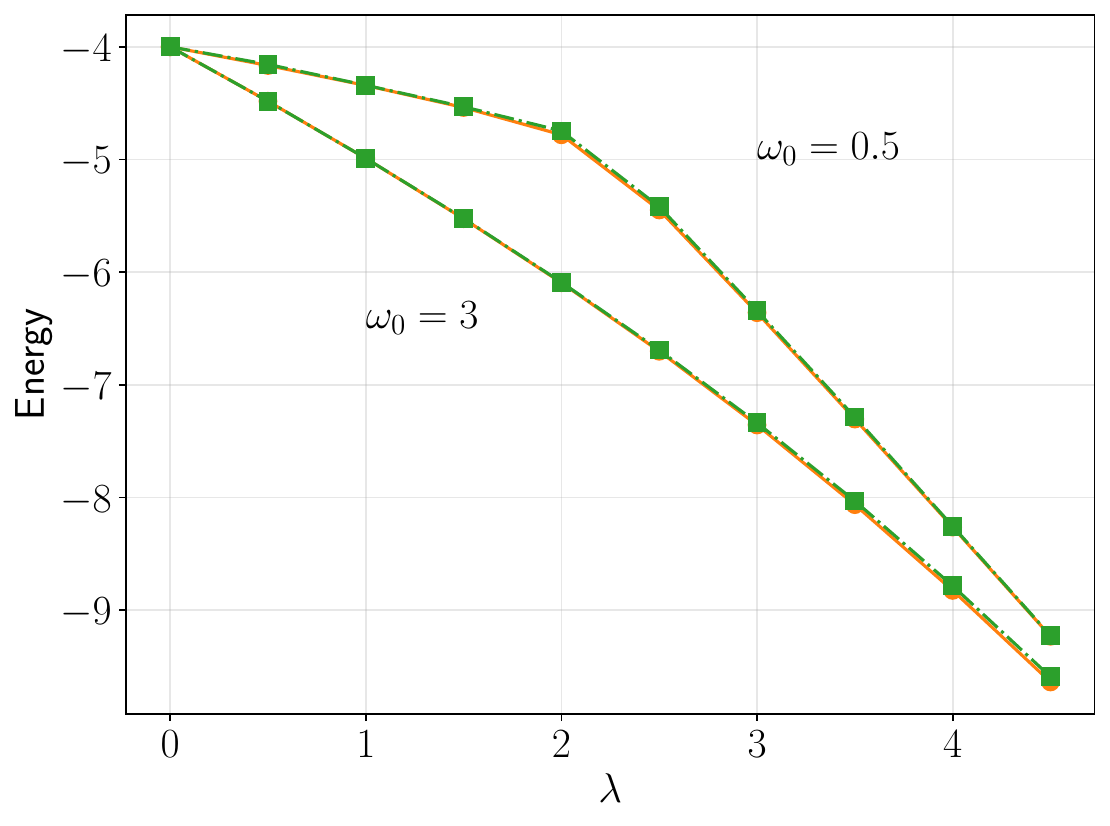}
   }\qquad
   \subfloat[Bond]{
      \includegraphics[width=0.9\columnwidth]{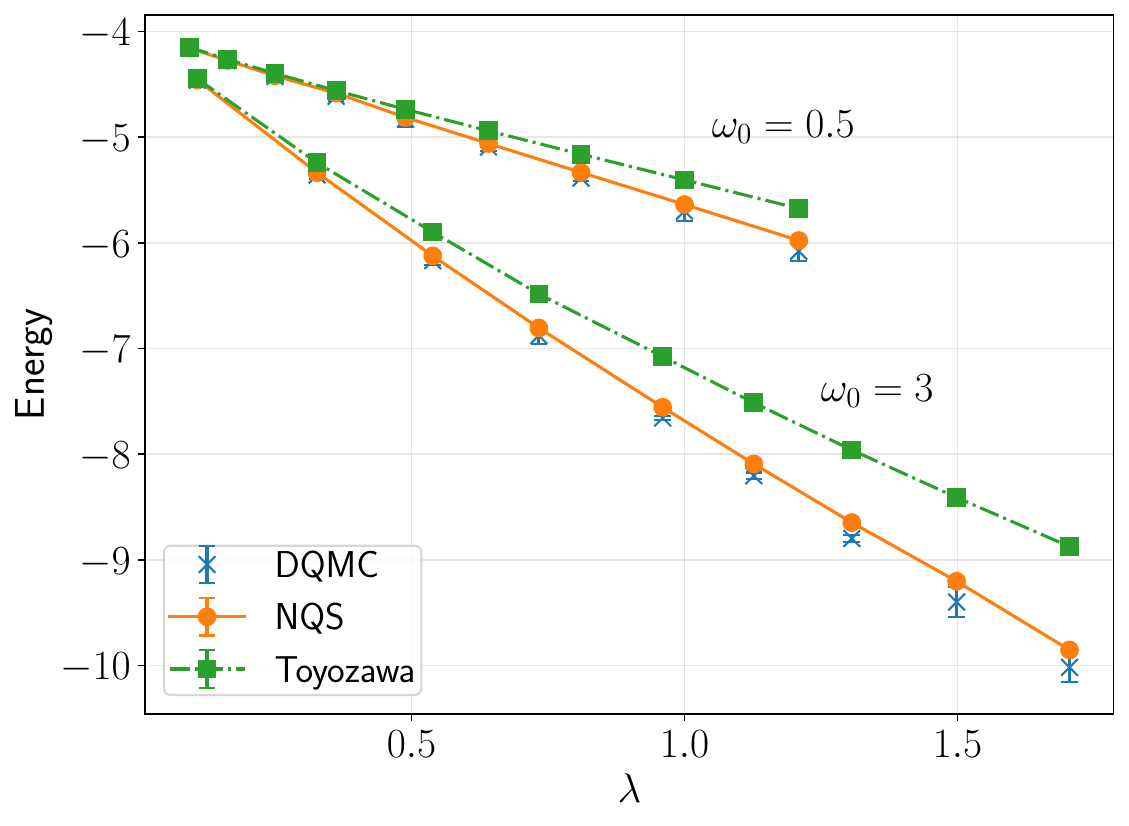}
   }\qquad
   \subfloat[SSH]{
      \includegraphics[width=0.9\columnwidth]{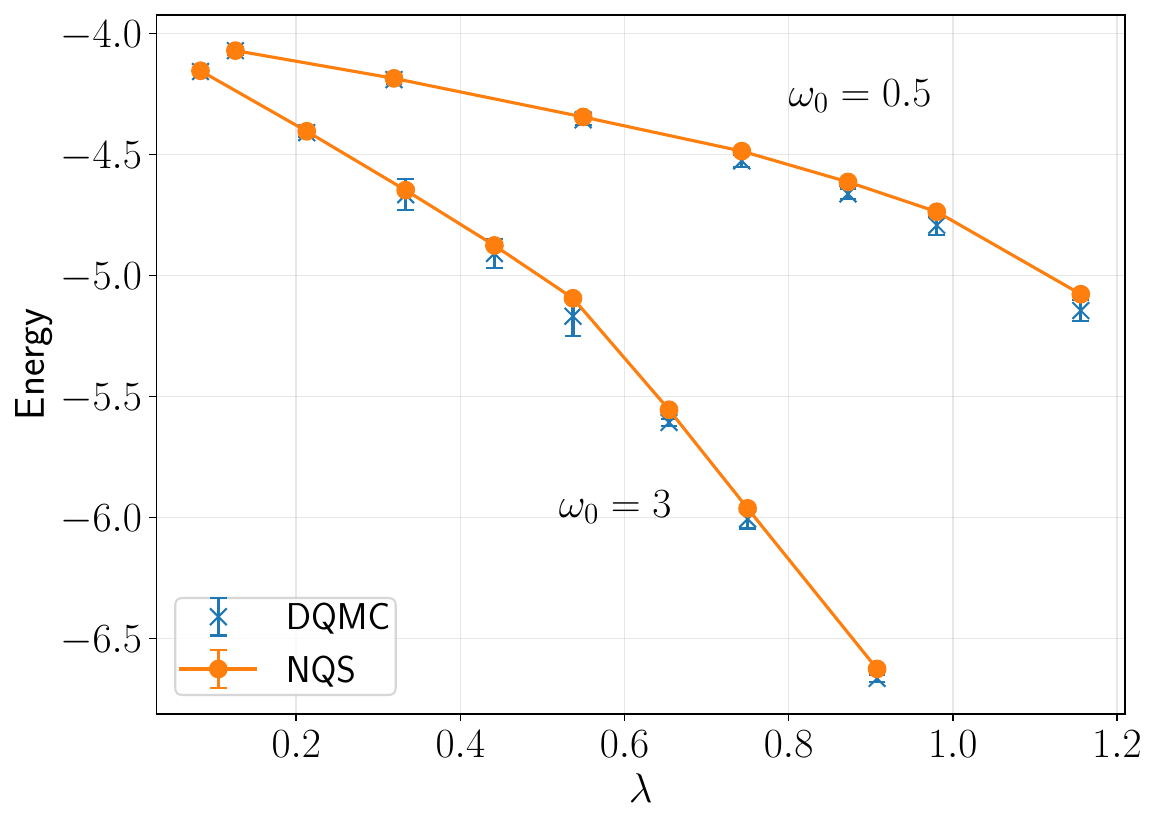}
   }
   \caption{Ground state energy of the Holstein, Bond, and SSH polarons on a two-dimensional lattice for \(\omega_0=0.5\) (upper curves) and \(\omega_0=3\) (lower curves). DQMC energies taken from Ref. \citenum{zhang2021peierls}}\label{fig:2d_polarons}
\end{figure*}

\subsubsection{Dispersive phonons}\label{sec:dispersive_phonons}
To demonstrate the robustness of our optimization protocol, we present a calculation on the ground state band of a modified Holstein model polaron with dispersive phonons. This system was studied using variational exact diagonalization (VED) in Ref. \citenum{bonvca2021dynamic}. We use the phonon dispersion given by
\begin{equation}
   \omega(q) = \omega_0 + 2t_{\text{ph}}\cos(q),
\end{equation}
where \(t_{\text{ph}}\) is the phonon hopping amplitude. The remaining parts of the Hamiltonian are identical to the usual Holstein model. We consider the parameters \(t_{\text{ph}}=0.4\) and \(\lambda=\frac{g^2}{2t_{\text{el}} \sqrt{\omega_0^2 - 4t_{\text{ph}}^2}}=0.5\). The phonon dispersion in this case bends downward due to the negative hopping amplitude, and results in a peculiar polaron band structure due to multiple avoided crossings with multiple phonon excitations. Fig. \ref{fig:ph_dispersion} shows the ground and first excited state bands obtained using a momentum space NQS on a 42 site chain. We find the ground state NQS energy to be in excellent agreement with VED results. The VMC optimization remarkably found the correct ground state at all \(k\) points starting with a completely random initial guess and did not encounter problems with trapping in local minima. The Toyozawa wave function is very accurate close to \(k=0\), but its error is substantial at larger \(k\) values. The NQS first excited state energy, obtained using LR-VMC (discussed in detail in Sec. \ref{sec:dynamical_properties}), is also in good agreement with VED results. We found some discrepancies at intermediate \(k\) values between the first two avoided crossings, which we attribute to finite-size effects. The inset shows the convergence of the first excited state energy at \(k=\frac{\pi}{3}\) with respect to the size of the lattice, showing a slow convergence with the lattice size. 

\begin{figure*}[t]
   \centering
   \subfloat[Holstein]{
      \includegraphics[width=0.9\columnwidth]{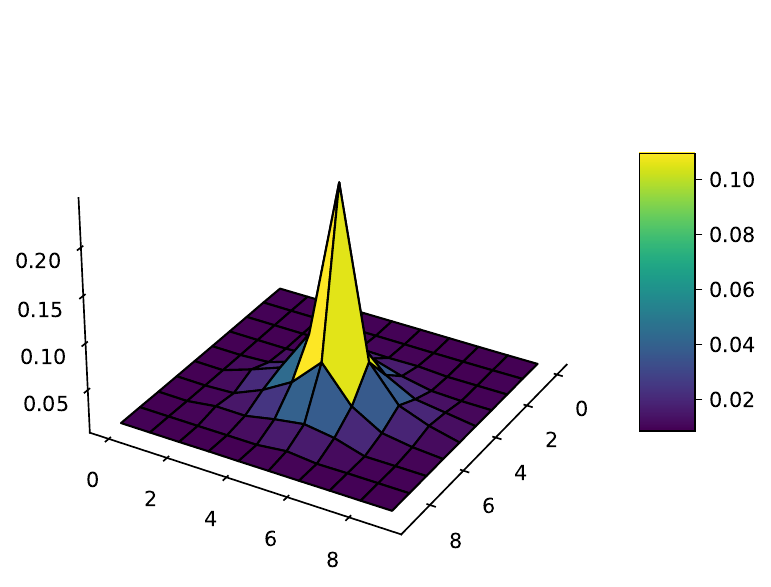}
   }\qquad
   \subfloat[Bond]{
      \includegraphics[width=0.9\columnwidth]{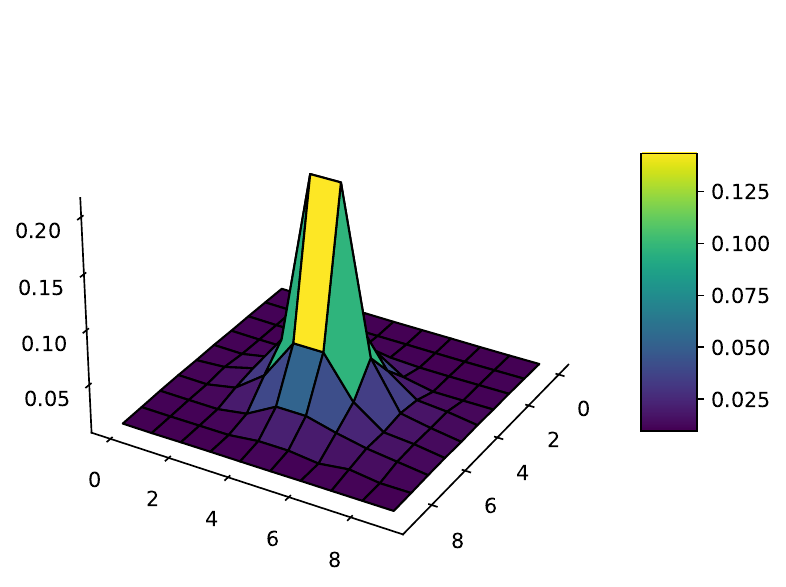}
   }
   \caption{Electron phonon correlation functions for the Holstein and Bond models with  \(\omega_0=1\) and \(\lambda=0.5\) calculated on a \(10\times 10\) lattice. For the Bond model, we show the correlation function with phonon modes on the bonds in the \(x\) direction.}\label{fig:correlation_functions}
\end{figure*}

\subsubsection{Two-dimensional polarons}\label{sec:2d_polarons}
Fig. \ref{fig:2d_polarons} illustrates the behavior of the polaron ground state energy on two-dimensional lattices. We used a \(10\times 10\) lattice for all calculations, which was confirmed to be sufficiently large to reach convergence within stochastic error bars for all cases presented here. For the Holstein model, similar to the one-dimensional chain considered previously, we find that the ground state energy is readily converged with a small number of hidden neurons. In fact, the Toyozawa wave function (equivalent to one hidden neuron) is very accurate in this case even on the two-dimensional lattice. We see the self-trapping crossover in the Holstein model clearly in this plot. The Bond and SSH models are considerably more difficult to solve for the NQS approach indicating the complexity of the off-diagonal coupling. We compare our energies to DQMC results reported in Ref. \citenum{zhang2021peierls}. For the Bond model, we are able to converge NQS energies to the DQMC results with a larger number of hidden neurons, around 100 for the largest couplings. The Toyozawa ansatz exhibits substantial errors that increase with the size of the coupling. The SSH model at large couplings has its ground state at nonzero \(k\) due to the negative next-nearest neighbor hopping amplitude induced by eph coupling. This behavior has been attributed to the unphysical linear nature of the interaction at strong couplings and was shown to disappear with more realistic nonlinear couplings.\cite{prokof2022phonon} Nevertheless, we find that the NQS is able to capture this shift of the ground state off the band center accurately. In this regime of stronger coupling, optimization becomes very slow and the ansatz requires a large number of hidden neurons to reach convergence. In these cases, we extrapolate our energies with respect to the number of hidden neurons (see Appendix \ref{app:variance_extrapolation}). We do not show the Toyozawa ansatz energies in this case because we were unable to converge them reliably. 

We are also able to calculate ground state properties other than energy using NQS. Fig. \ref{fig:correlation_functions} shows the electron-phonon correlation functions for the two-dimensional Holstein and Bond models. The correlation function is defined as
\begin{equation}
   \xi(|i-j|) = \frac{\langle n_i x_j\rangle}{\langle n_i\rangle},
\end{equation}
where \(n_i\) and \(x_j\) denote electron occupation and phonon displacement, respectively. The correlation function is a measure of the spatial extent of the polaron. The difference between the diagonal and off-diagonal couplings is immediately apparent from the correlation functions. In the Holstein model, the phonon cloud builds around the electron at a site, whereas in the Bond model, the electron hops between two sites exciting the phonons in surrounding bonds. We note that only the correlation function is localized in polaron models, but the exact ground state remains delocalized reflecting the lack of a true self-trapping transition in these models.

\begin{figure}[htp]
   \centering
   \includegraphics[width=0.5\textwidth]{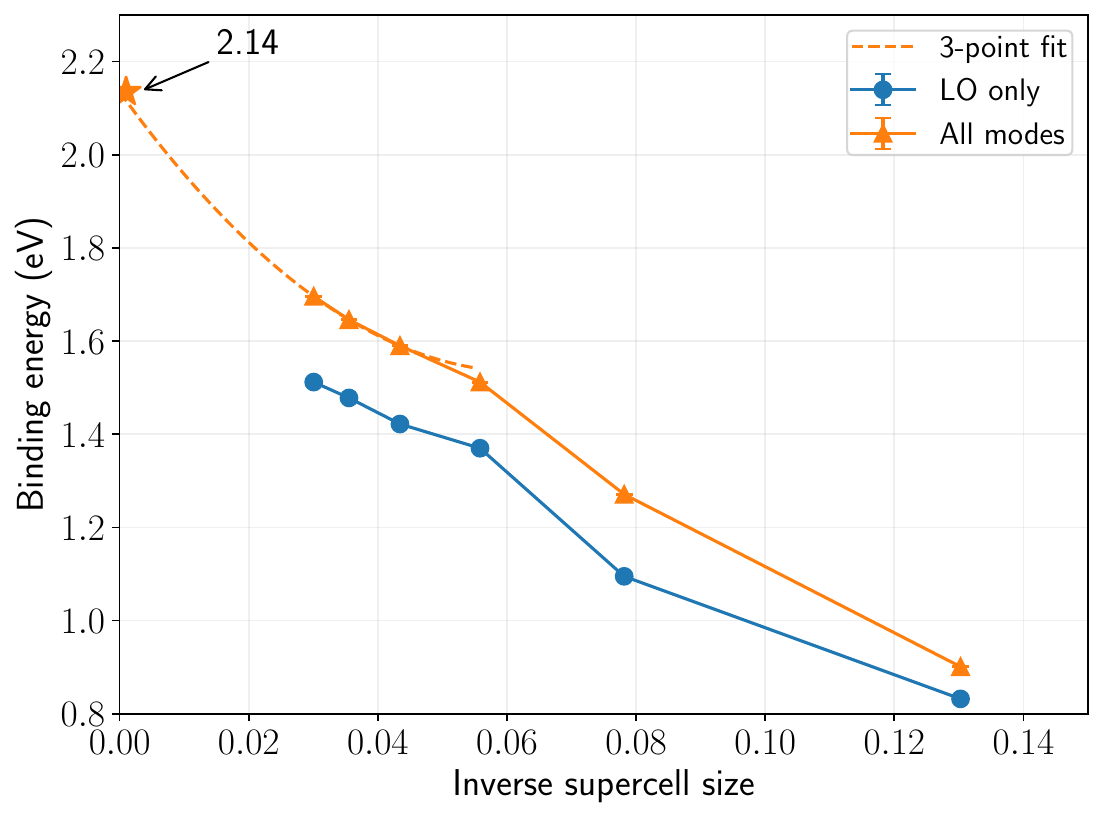}
   \caption{Binding energy of the LiF hole polaron obtained using NQS. ``LO only" indicates calculations performed with only the longitudinal optical phonon mode, whereas the ``All modes" energies are obtained with all three acoustic and optical modes. The star indicates the binding energy extrapolated to the thermodynamic limit using a quadratic fit.}\label{fig:lif_hole_be}
\end{figure}

\begin{figure}[htp]
   \centering
   \includegraphics[width=0.5\textwidth]{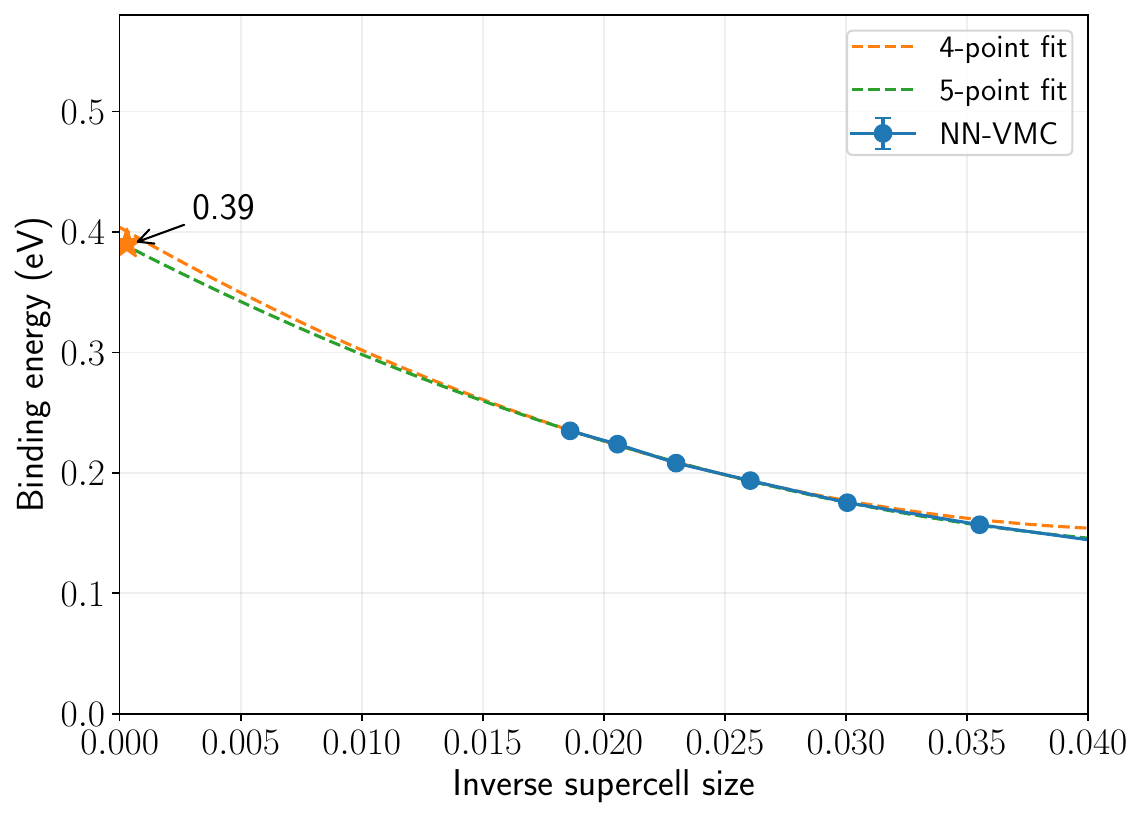}
   \caption{Binding energy of the LiF electron polaron obtained using NQS. The star indicates the binding energy extrapolated to the thermodynamic limit using a quadratic fit. Agreement between the fits using the largest 4 and 5 grids simulated demonstrate the fidelity of this fitting procedure.}\label{fig:lif_elec_be}
 \end{figure}

\begin{figure*}[t]
   \centering
   \subfloat{
      \includegraphics[width=0.9\columnwidth]{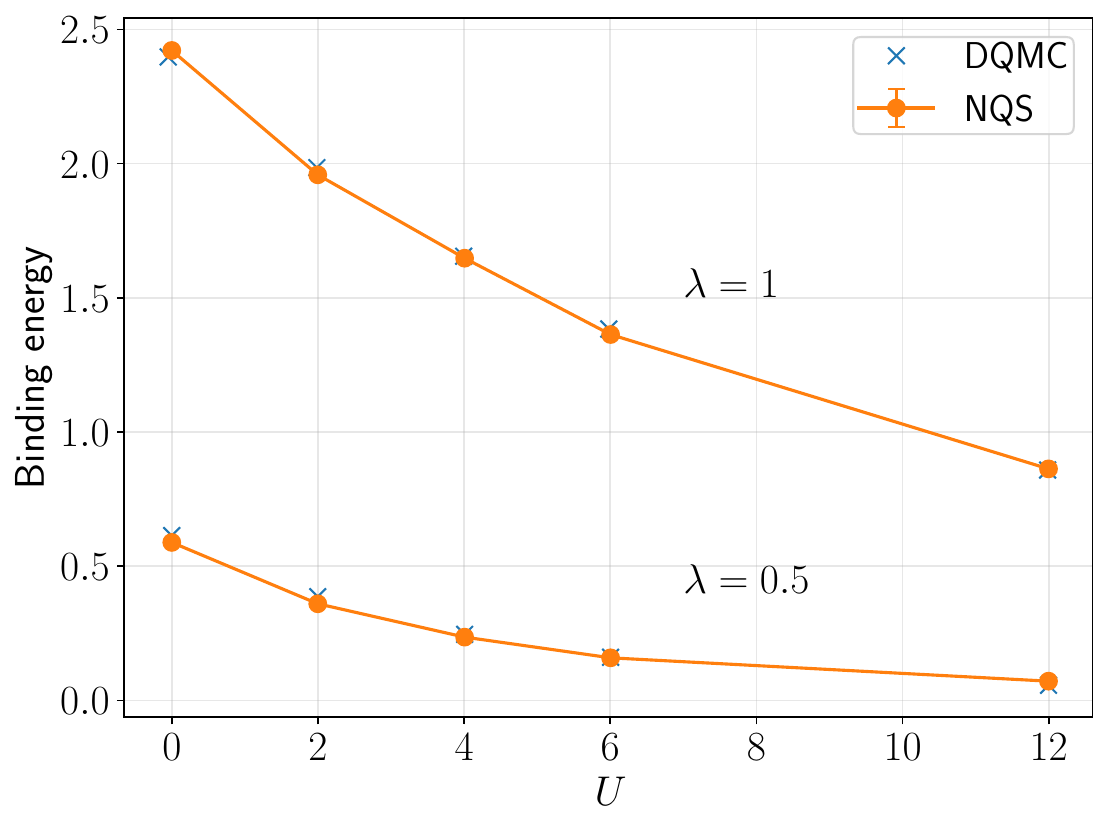}
   }\qquad
   \subfloat{
      \includegraphics[width=0.9\columnwidth]{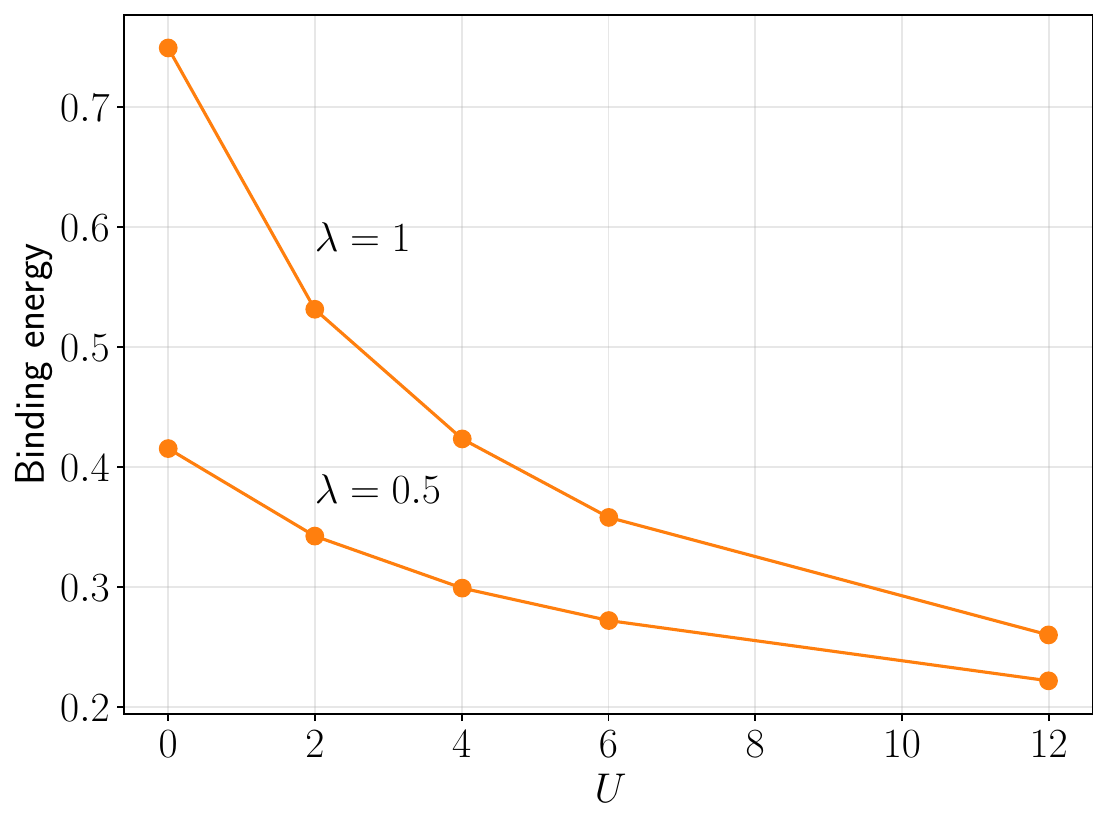}
   }
   \caption{Binding energy of the Bond and SSH bipolarons on a two-dimensional square lattice with \(\omega_0=0.5\) and \(\lambda=0.5\) (lower curve) and 1 (upper curve) as a function of on-site repulsion \(U\). DQMC results for the Bond model taken from Ref. \citenum{zhang2022bond}.}\label{fig:bipolarons}
\end{figure*}

\subsubsection{\textit{Ab initio} calculations of polaron binding energies in LiF}\label{sec:ab_initio}
For the final polaron example, we calculate the hole and electron polaron binding energies in lithium fluoride (LiF) from first principles.\cite{sio2019ab,lee2021facile} This is a polar ionic crystal with a large band gap (14.2 eV experimental optical gap, 8.9 eV calculated PBE KS gap).\cite{sio2019ab} We obtain the Hamiltonian parameters from DFT calculations using Quantum Espresso\cite{Giannozzi_2017} with the EPW package.\cite{ponce2016epw} We include three valence bands arising mainly from the \(p\) orbitals of fluorine in our hole polaron calculation, while one conduction band was included in the electron polaron calculation. Higher energy bands may in principle be included, but are omitted here. Previous work has indicated that these bands provide the dominant contribution to the binding energy,~\cite{sio2019ab,lee2021facile} however inclusion of more remote bands may provide quantitative contributions the polaron energetics.  The relevant bands are coupled to three optical and three acoustic mode vibrational phonon branches. Due to the polar nature of this system, the hole coupling to the longitudinal optical (LO) mode is the strongest and largely of the Fr\"{o}hlich type. 

Due to the very strong coupling of the hole with phonons, we expect the hole polaron ground state to be readily describable with a strong coupling ansatz. This is borne out in our calculations shown in Fig. \ref{fig:lif_hole_be}. We calculate ground state energies on progressively larger \(k\) point grids ranging from \(3\times 3\times 3\) to \(13 \times 13 \times 13\). Due to the large number of input sites in the NQS state, we are restricted in the number of hidden neurons we can employ in this calculation. Therefore as a validation of our ansatz, we also calculate energies on a smaller model including just the LO phonon mode in the Hamiltonian, which contributes the bulk of the binding energy. We find that increasing the number of hidden neurons provides modest improvements (of the order of only 10 meV) over the Toyozawa ansatz as expected in a strong density coupling case. Using a quadratic fit of the energies obtained for the three largest grids we obtain a binding energy of 2.14 eV in the thermodynamic limit. Given the uncertainties in the extrapolation procedures due to computational limitations of the grid sizes employed, it is satisfying that our value of the binding energy is in reasonable agreement with strong coupling perturbation theory calculations (1.96 eV) and results obtained from a novel all-coupling wave function ansatz (1.94 eV),\cite{robinson2024abinitio} as well as a calculation using many-body Green's function theory (2.20 eV).\cite{lafuente2022unified} 

The electron polaron is weakly coupled and more delocalized, requiring larger \(k\)-grids to converge the binding energy. We performed NN-VMC calculations on grids of sizes up to \(21\times 21\times 21\). Fig. \ref{fig:lif_elec_be} shows the binding energy of the electron polaron as a function of the inverse grid size. We used a quadratic fit to extrapolate to the thermodynamic limit in this case as we found it to model the data better. The quadratic fits to the largest 4 and 5 grids produce similar extrapolated energies within 10 meV of each other. The five-point estimate of 0.39 eV is lower than the 0.6 eV obtained using a Green's function method in Ref. \citenum{lafuente2022unified}, possibly due to the inclusion of contributions from higher conduction bands and quadratic coupling in Ref. \citenum{lafuente2022unified}, or simply due to the fact that the approach of Ref. \citenum{lafuente2022unified} is approximate.    

The largest uncertainty in these binding energies, which are not reported, arises simply due to the nature of the extrapolation from small grid sizes. For example, in the case of the hole polaron, a linear extrapolation using the two largest grids leads to an extrapolated binding energy of 1.97 eV compared to the 2.14 eV obtained using a quadratic fit. A future study will be devoted to a more careful extrapolation of these values. With improvements in our numerical methodology, including the use of locality of interactions and low-rank compression,\cite{luo2024data} we expect to be able to perform calculations on even larger grids in the future. 

\subsubsection{Bond and SSH model bipolarons}\label{sec:bipolarons}
We now turn to the calculation of two interacting electrons coupled to phonons. In cases of strong eph interactions, one can obtain a bound state of electrons, termed a bipolaron, due to the phonon-mediated attraction overcoming the Coulomb repulsion. Recent work on bipolarons in the Bond and SSH models shows that light yet mobile bipolarons exist in these models even with strong Coulomb interactions. In Fig. \ref{fig:bipolarons}, we compare ground state energies obtained using NQS states with DQMC values reported in Ref. \citenum{zhang2022bond} for the Bond model. These calculations were performed on a \(12\times 12\) lattice. Note that there is some cancellation of finite size errors due to attractive eph and repulsive e-e interactions. The largest NQS state here used 150 hidden neurons. For the SSH model, we performed extrapolations with respect to the number of hidden neurons to obtain the reported energies. No calculations of the binding energy have been reported for this model in the literature. However, they display a trend similar to the Bond model with the bipolarons staying bound even at large Coulomb interactions. We have performed calculations for the one-dimensional case to compare against the exact diagonalization results presented in Ref. \citenum{sous2018light} and found very close agreement, providing further validation for the accuracy of the results for the square lattice in lieu of benchmark results.

\begin{figure}[htp]
   \centering
   \includegraphics[width=0.5\textwidth]{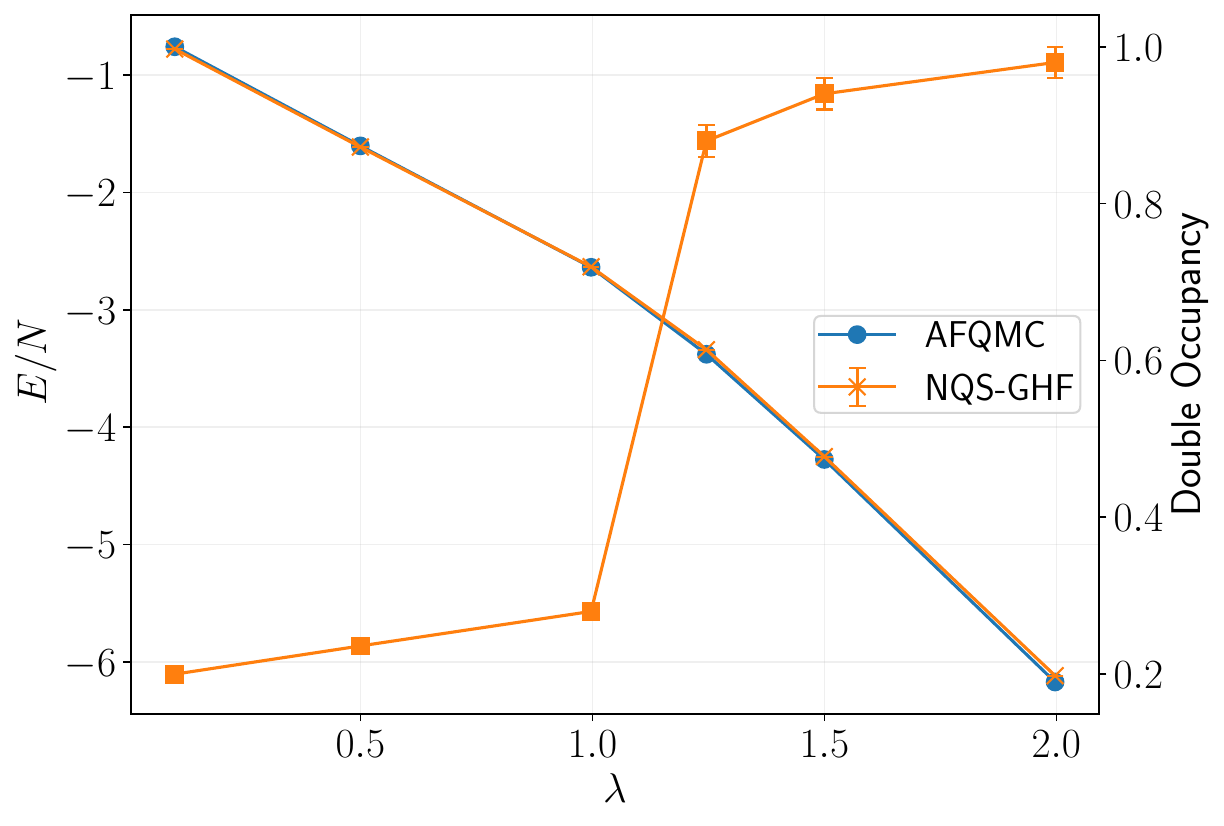}
   \caption{Ground state energy of the Holstein-Hubbard model at half-filling on a 20 site chain with \(U=4\) and \(\omega_0=1\) (corresponding to the axis on the left). Square symbols show double occupancy in the NQS state (corresponding to the axis on the right). AFQMC results taken from Ref. \citenum{lee2021constrained} }\label{fig:hh}
\end{figure}

\subsubsection{Holstein model with finite filling}\label{sec:holstein_n}  

\begin{table}
   \centering
   \caption{Ground state energy per site of the Holstein model at half-filling \(\omega_0=1\) and \(g=1.5\). Jastrow and RBM values are taken from previous VMC studies.}\label{tab:holstein_n}
   \begin{tabular}{ccccccccc}
      \hline
      \(N_{\text{sites}}\) &~& Jastrow\cite{ohgoe2014variational} &~& RBM\cite{nomura2020machine} &~& MLP (this work) &~& GFMC\cite{mckenzie1996quantum}\\
      \hline
      4 && -0.8904(5) && - && -0.89394(7) && -0.895(1) \\
      6 && -0.8583(3)  && -0.868 && -0.8674(2) && -0.868(1) \\
      8 && -0.8424(3) && -0.861 && -0.8606(1) && -0.861(2) \\
      16 && -0.8388(5) && -0.855 && -0.8535(2) && -0.854(1) \\
      \hline
   \end{tabular}
\end{table}

We consider a Holstein model with spinless fermions that do not interact directly but only do so indirectly through the eph interaction. Previous QMC studies, including one using RBM NQS, used this model at half-filling on a one-dimensional chain as a benchmark, with Ref. \citenum{mckenzie1996quantum} providing numerically exact Green's function Monte Carlo (GFMC) ground state energies. The form of the Hamiltonian used is slightly different from the Holstein model shown in Table \ref{tab:models}, with the eph coupling term given by
\begin{equation}
   H_{\text{eph}} = g\sum_i \left(c^{\dagger}_ic_i-\frac{1}{2}\right)(b^{\dagger}_i+b_i). 
\end{equation}
We use periodic boundary conditions for systems with an odd number of fermions and antiperiodic boundary conditions otherwise, in keeping with previous studies. 

Table \ref{tab:holstein_n} shows the ground state energy per site for different lattice sizes for \(\omega_0 = 1\) and \(g=1.5\). Ref. \citenum{ohgoe2014variational} used electron-phonon and conventional electron-electron Jastrow factors on top of a BCS reference wave function within VMC, whereas Ref. \citenum{nomura2020machine} used an RBM Jastrow on top of the same reference. The correlating factors in both these studies are non-negative and thus inherit their sign structure from the reference state. In contrast, we have used a complex NN correlator (Eq. \ref{eq:nn_wave_function}) on top of a simple Hartree-Fock reference state. We impose translational invariance on this wave function by explicitly symmetrizing overlaps. This represents a more flexible ansatz with a general sign structure, but also poses a challenging problem, requiring the NN to capture the relevant physics. The difficulties in learning sign structures of many-body wave functions using NQS have been reported in the literature.\cite{westerhout2020generalization} Despite these challenges, we find that our method is able to obtain ground state energies in good agreement with the numerically exact GFMC results. We found that MLPs with up to \(10N_{\text{sites}}\) hidden neurons were sufficient to converge the results to within the stochastic error bars of the GFMC energies. They represent a significant improvement over the conventional Jastrow results of Ref. \citenum{ohgoe2014variational} and are comparable to the converged RBM results of Ref. \citenum{nomura2020machine} (which were estimated from Fig. 4 of this paper).

\subsubsection{Hubbard-Holstein model}\label{sec:hh}
In this section, we present results for the Hubbard-Holstein model on a one-dimensional chain of 20 sites at half-filling with \(U=4\). These calculations were performed in the site basis. We used a generalized Hartree Fock (GHF) state as the reference antisymmetric electronic state multiplied with an NQS eph Jastrow factor,

\begin{equation}
   \inner{\mathbf{n}}{\psi} = \inner{\mathbf{n}}{\psi_{\text{NQS}}}\inner{\mathbf{n}_e}{\psi_{\text{GHF}}},
\end{equation}
where \(\mathbf{n}_e\) denotes just the electronic part of the input configuration. Note that the GHF state breaks the spin projection \(S_z\) symmetry, which is restored by the VMC sampling procedure. We use a real MLP with 40 hidden neurons as the Jastrow factor. The sign structure is therefore inherited from the reference GHF state, which is a bias in this calculation. Despite these considerations, calculations on the half-filled pure Hubbard model indicate that this state is an excellent approximation for describing electronic correlation. In this model, as the eph coupling is increased, the system undergoes a transition from a quasiordered Mott insulator phase to a charge density wave (CDW) phase. This is reflected in the energies shown in Fig. \ref{fig:hh}. Agreement with AFQMC results, obtained using the constrained path approximation,\cite{lee2021constrained} is seen to be very good. We note that to converge the wave function to the correct CDW minimum at larger couplings, we perform the initial GHF calculation with an effective attractive coupling given by \(U_{\text{eff}} = U - 4 \lambda\). We see that the electronic double occupancy of the lattice sites given by

\begin{equation}
   d = \frac{2}{N}\sum_i \langle n_{i\uparrow}n_{i\downarrow}\rangle,
\end{equation}
changes rapidly near the transition point. Double occupancy close to one is an indicator of electron pairing in the CDW phase.

\subsection{Dynamical properties}\label{sec:dynamical_properties}
One particle spectral functions of polaron models have been extensively studied in the past and serve as a good benchmark for the current method. Here we will work in the momentum space basis to calculate the one-particle spectral function at zero temperature given by
\begin{equation}
   A(k,\omega) = -\frac{1}{\pi}\text{Im}\langle 0| c_k \frac{1}{\omega - H + i\eta} c^{\dagger}_k|0\rangle,
\end{equation}
where \(\ket{0}\) denotes the vacuum state. This quantity is directly related to angle-resolved photoemission spectroscopy (ARPES) measurements.\cite{sobota2021angle} Accurate numerical calculations have been performed using variations of exact diagonalization methods,\cite{marsiglio1993spectral,zhang1999dynamical,weisse2006kernel} DMRG,\cite{jansen2020finite} hierarchical equation of motion,\cite{mitric2022spectral} and generalized Green's function cluster expansion\cite{carbone2021numerically} methods on modestly sized systems and mostly for the Holstein model. While it may be difficult or impractical to do so in certain Hamiltonian parameter regimes, these methods nonetheless have the virtue of allowing the user to assess the convergence of the results obtained to the exact limit.

\begin{figure}[htp]
   \centering
   \includegraphics[width=\columnwidth]{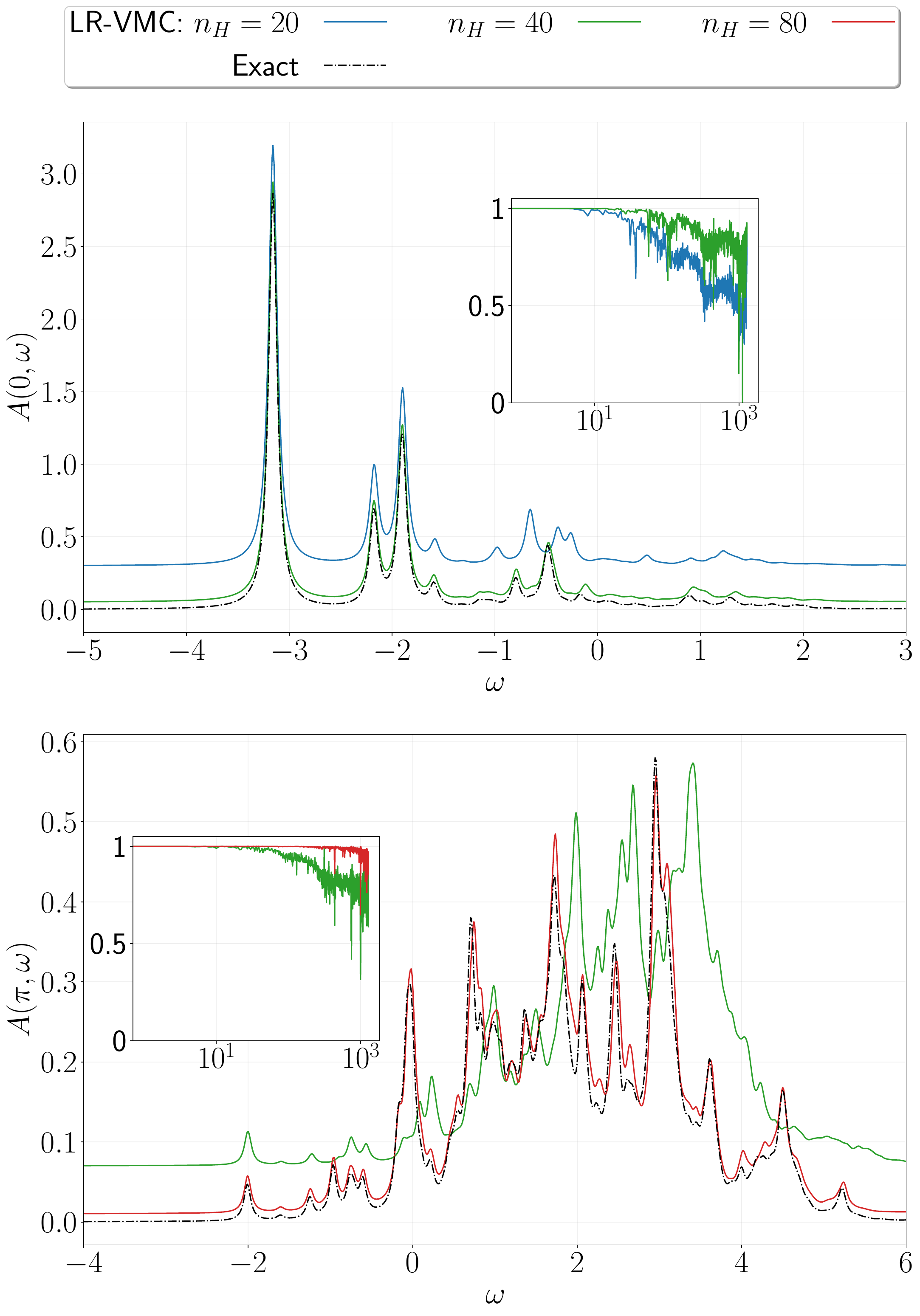}
   \caption{Convergence of the Bond model (8 sites, \(\omega_0=1\), \(\lambda=1\)) polaron spectral functions at \(k=0\) (top) and \(k=\pi\) (bottom). LR-VMC spectral functions (\(\eta=0.05\)) with different numbers of hidden neurons (\(n_H\)) in the ansatz are compared to the exact spectral function in a truncated space containing a maximum of 5 phonons. The LR-VMC spectral functions have been shifted up for clarity. Insets show the norm of the projection of the lowest \(10^3\) exact energy eigenstates on the LR-VMC tangent space, \(\| P_{\text{LR}} \ket{E_i}\|\).} \label{fig:spec_convergence}
\end{figure}

\subsubsection{Convergence of the LR-VMC method}\label{sec:lr_vmc_convergence}
As an illustrative example, we compare our results with exact diagonalization of the Bond model (\(\omega_0=1\), \(\lambda=1\)) polaron in an 8 site chain in a truncated space with a maximum of 5 phonons. We restricted the number of phonons in VMC sampling to the same number for consistency. To focus solely on the quality of the LR approximation and its convergence with the quality of the ansatz, we obtained the LR-VMC results deterministically in this small example by simply summing over all the eph configurations instead of VMC sampling. Discussion of the biases due to sampling can be found in Appendix \ref{app:lr_vmc_sampling}. Fig. \ref{fig:spec_convergence} shows the comparison of the exact spectral function with an LR-VMC calculation for different numbers of hidden neurons in a single hidden layer MLP state. \(n_H\) in the figure denotes the number of hidden neurons in the magnitude and phase NNs each. 

\begin{figure*}[htp]
   \centering
   \includegraphics[width=\textwidth]{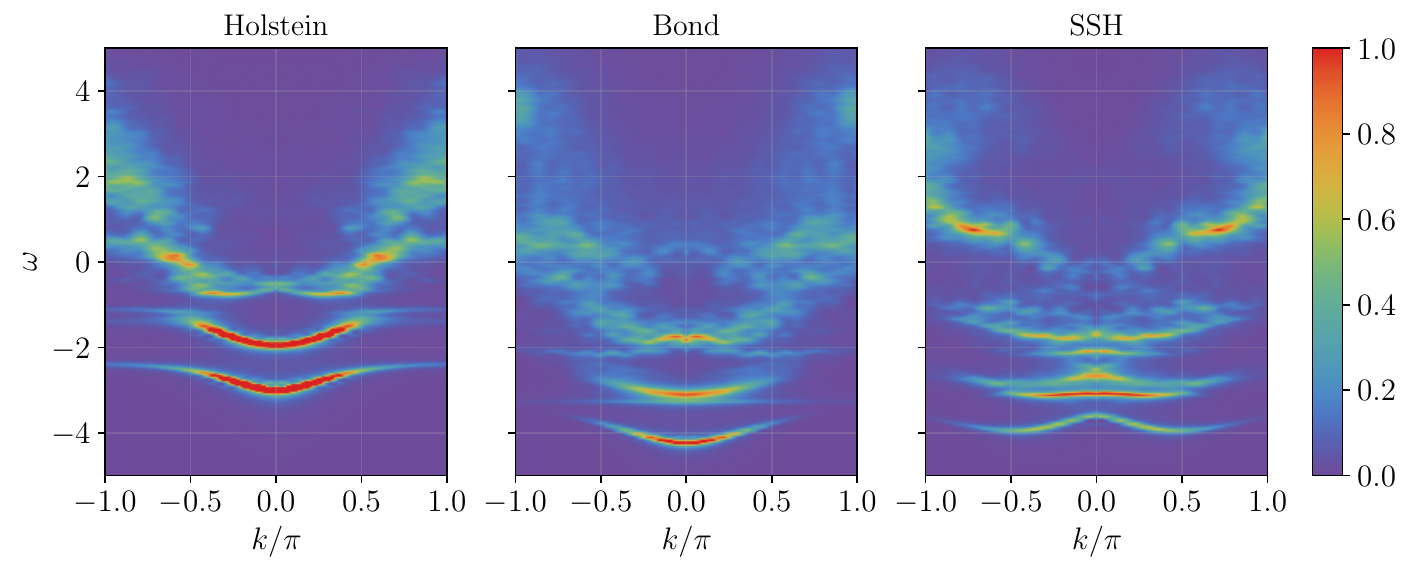}
   \caption{LR-VMC spectral functions for Holstein, Bond and SSH model polarons on a 30 site chain. \(\omega_0=1\) and \(\lambda=1\) for all models. Interpolated from data obtained at 30 \(k\)-points with a Lorentzian broadening \(\eta=0.05\).}\label{fig:spec_map}
\end{figure*}

At \(k=0\), the dominant quasiparticle peak corresponds to the polaron ground state. Subsequent peaks result from the addition of phonons to the polaron, the first one being at roughly \(\omega_0\) energy above the ground state, marking the onset of the phonon excitation continuum corresponding to states of the polaron with additional phonon excitations. LR-VMC on top of the \(n_H=20\) state captures the first 4 peaks very well but starts to deviate from the reference spectral function at higher energies. Increasing the number of hidden neurons in the ansatz to 40, we see that the agreement with the reference improves at higher energies. The norm of the projection of exact energy eigenstates on the LR-VMC tangent space, \(\| P_{\text{LR}} \ket{E_i}\|\), shown in the inset, is a measure of the quality of the LR approximation. We see that the overlap decays for higher energy states, but lower energy states are represented remarkably well even with \(n_H=20\). This evidence suggests that the tangent space states likely represent simple excitations on top of the ground state, like those present in the low-lying eigenstates. 

At the band edge, \(k=\pi\), the spectrum is concentrated in the higher energy regions corresponding to the incoherent phonon continuum. The low energy states in this case have the electron close to the band minimum with high momentum phonons leading to a low spectral weight. The \(n_H=40\) wave function captures this low part of the spectrum well as seen from the projection norm shown in the inset. Because of the nature of the LR ansatz, it takes many more hidden neurons, \(n_H=80\), to nearly converge to the exact spectral function at higher energies. The tangent space for this state essentially represents the whole truncated space as evidenced by the norms of the energy eigenstate projections. We note that the more compact wave function still produces a qualitatively correct structure in the incoherent region. 

\subsubsection{Spectral functions of polaron models}\label{sec:polaron_spectra}
In Fig. \ref{fig:spec_map}, we show the spectral functions of the Holstein, Bond, and SSH model polaron on a 30 site chain. The spectral functions were converged with respect to the number of hidden neurons in the ansatz up to the stochastic error. For \(\omega_0=1\) and \(\lambda=2\) used here, we verified that finite size effects are negligible. In the Holstein model spectral function, the first excited state at \(k=0\) is exactly \(\omega_0\) in energy above the ground state, indicating an unbound state of a phonon well separated from the ground state polaron. We also note the appearance of a discrete state immediately above the one phonon continuum band more prominent near the band edge. In Ref. \citenum{vidmar2010emergence}, this was characterized as an antibound state between the polaron and a phonon which has vanishing spectral weight at \(k=0\). We do not see the nondispersive repulsive state seen in that work, which they attributed to the finite size of the lattice used in that work. 

The Bond model has a higher binding energy compared to the Holstein model while having a lighter polaron mass at the same time. This behavior which can be attributed to the coupling of phonons to carrier hopping terms has been noted in previous work.\cite{carbone2021bond} We also note the presence of a bound first excited state just below the one phonon continuum band. While it is almost nondispersive at this coupling, it has a concave dispersion at stronger couplings (not shown here). This is in contrast to the Holstein model, where the first excited bound state at intermediate couplings has the same dispersion shape as the ground state. The SSH model has a very different spectrum compared to the other two models with the ground state at a non-zero momentum. We again see the appearance of a bound excited state below the first phonon continuum carrying a large spectral weight. 

\begin{figure}
   \centering
   \includegraphics[width=\columnwidth]{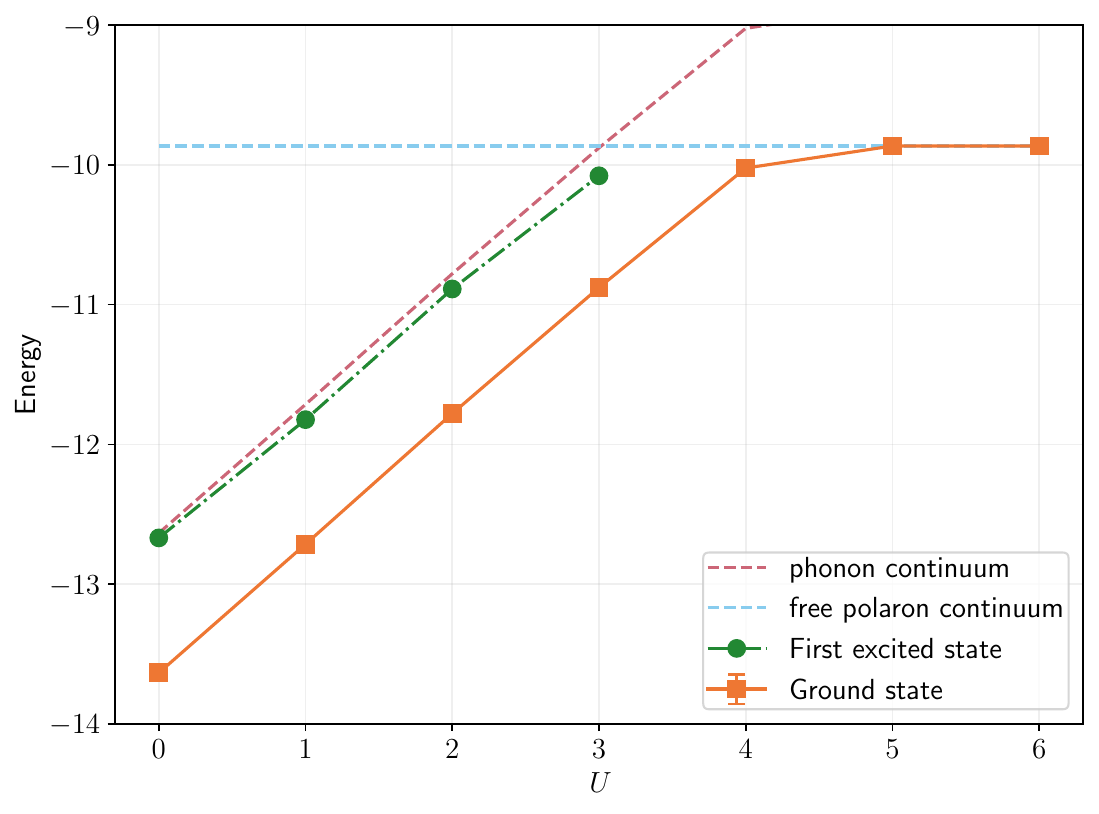}
   \caption{Low-lying bound states of the Holstein bipolaron (\(\omega_0=1\), \(\lambda=3.24\)) on a two-dimesnional lattice in the \(s\)-wave symmetry seetor. Dashed lines mark the onsets of one-phonon and free polaron continuua.}\label{fig:hol_bip}
\end{figure}

\subsubsection{Low-lying states of the Holstein bipolaron}\label{sec:bipolaron_excited_states}
We also calculated the ground and first excited states of the two-dimensional Holstein bipolaron as a function of the electronic interaction \(U\). Results on a \(10\times 10\) lattice are shown in Fig. \ref{fig:hol_bip}. For moderate eph coupling, the ground state in this model evolves from a strongly bound \(S_0\) bipolaron, with both electrons mostly on the same site, to a weakly bound \(S_1\) bipolaron with the two electrons on neighboring sites. Our ground state energies are in good agreement with those reported in Ref. \citenum{macridin2004two}. In the on-site regime, the energy increases nearly linearly with \(U\). Strong coupling arguments\cite{macridin2004two} suggest the presence of two singlet excited \(S_1\) states below the phonon and free-polaron continuua for weak to intermediate \(U\). These states are bound due to the interplay of effective kinetic exchange interactions with eph coupling. One of these states has a \(d\)-wave symmetry, while the other has \(s\)-wave symmetry. Since spatial symmetry is projected in our calculations, LR-VMC on top of the \(s\)-wave symmetric ground state only captures excited states in this symmetry sector. We find this excited state just below the phonon continuum for small \(U\). With increasing \(U\) it starts mixing with the \(S_0\) ground state and after the crossover, it becomes the ground state. 

\begin{figure*}[t]
   \centering
   \subfloat[Heisenberg chain]{
   \includegraphics[width=0.9\columnwidth]{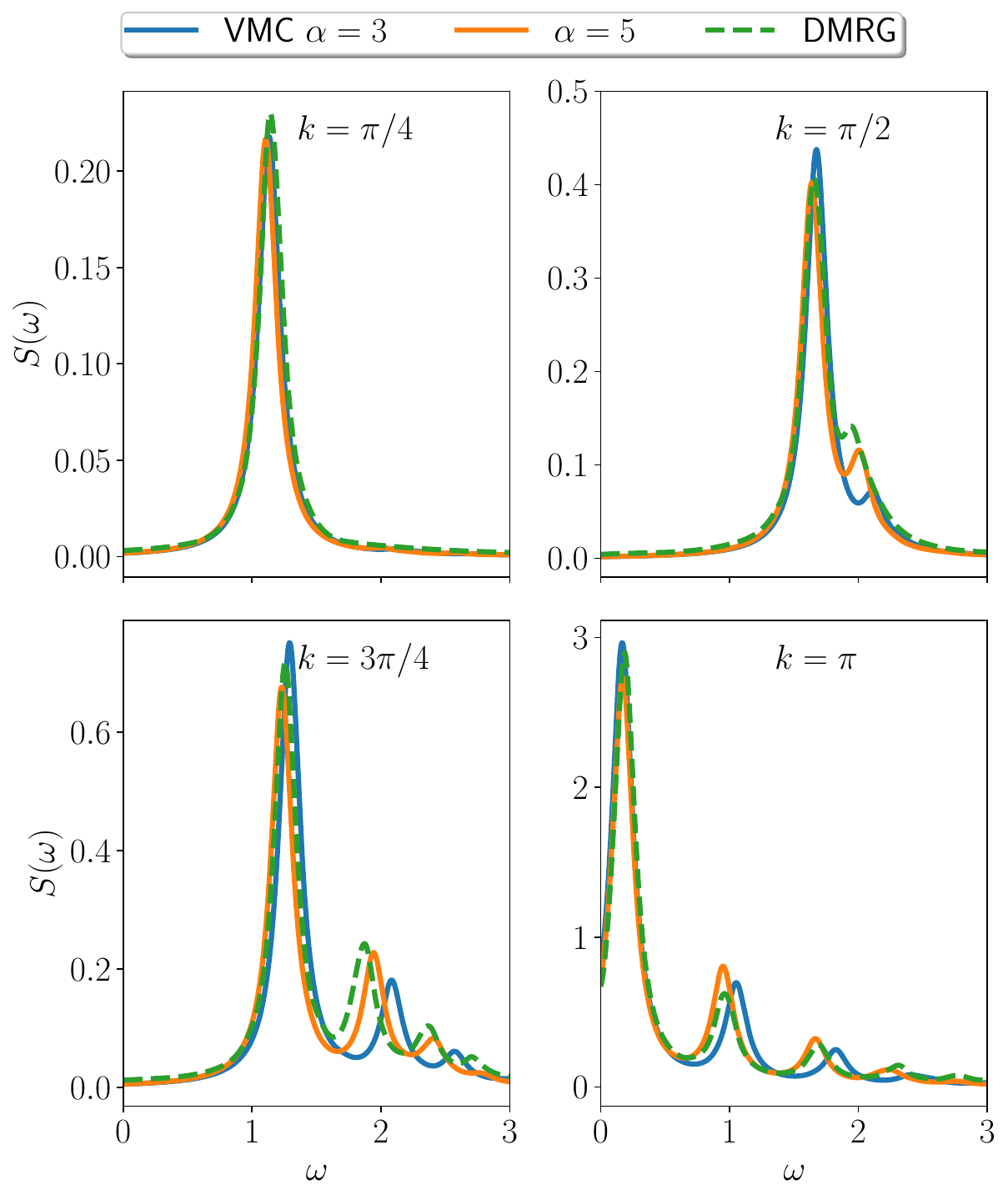}
   }\qquad
   \subfloat[Heisenberg-Bond 2D lattice]{
      \includegraphics[width=0.9\columnwidth]{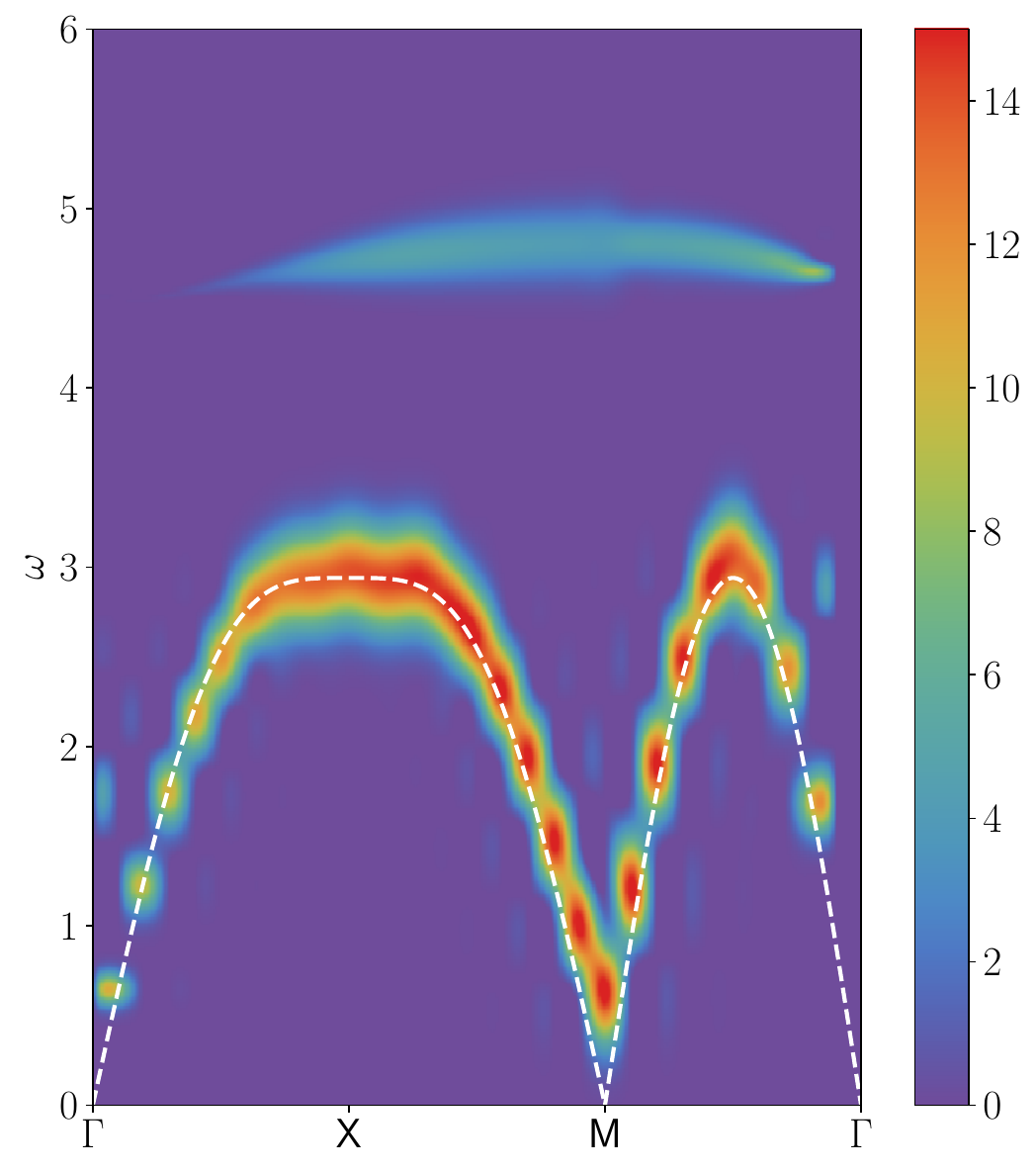}
   }
   \caption{(a) Convergence of the LR-VMC dynamical spin structure factor for the spin-1/2 Heisenberg model on a 32 site chain at different \(k\)-points with a broadening of \(\eta=0.1\). \(\alpha\) is the density of hidden neurons. (b) The same quantity for the Heisenberg-Bond model on a \(10\times 10\) square lattice with \(\omega_0=2\) and \(g=1\). The white dashed line shows linear spin wave magnon energies with a fitted \(J_{\text{eff}} = 1.47J\)}\label{fig:heisenberg_spec}
\end{figure*}

\subsubsection{Dynamical spin structure factor of the Heisenberg and Heisenberg-Bond models}\label{sec:heisenberg_spectra}

The Heisenberg model serves as a useful benchmark system for the calculation of dynamical properties since reference values can be obtained relatively easily for large systems using DMRG. This model has been used for assessing the performance of previous NQS based dynamical calculations, therefore also provides a good point of comparison. The spin-1/2 Heisenberg Hamiltonian is given by
\begin{equation}
   H = J\sum_{\langle ij\rangle} \mathbf{S}_i\cdot\mathbf{S}_{j},
\end{equation}
where \(\mathbf{S}_i\) are spin-1/2 operators on lattice sites and the exchange interactions are limited to nearest neighbors. We set \(J=1\) corresponding to an antiferromagnetic coupling. We use an NQS of the same form as Eq. \ref{eq:nn_wave_function} with the spin configuration used as the input. A quantity of interest in this model is the dynamical spin structure factor defined as
 \begin{equation}\label{eq:spin_structure_factor}
   \begin{split}
      S_{\sigma}(k,\omega) &= \frac{1}{N}\sum_{ij}e^{ik(i-j)}\int_{0}^{\infty}dt e^{i\omega t}\expecth{\psi_0}{S^z_i(t)S^z_j(0)}{\psi_0}\\
      &= -\frac{1}{\pi}\text{Im}\expecth{\psi_0}{S^z_{-k}\frac{1}{\omega-H+E_0+i\eta}S^z_k}{\psi_0},
   \end{split}
\end{equation}
where \(\ket{\psi_0}\) is the ground state wave function. Unlike the polaron spectral function calculations, here, we perform a single VMC optimization for the ground state at \(k=0\). The LR basis functions at all \(k\) points are obtained by momentum projection of the \(k=0\) LR space. 

We calculated \(S_{\sigma}(k, \omega)\) for the spin-1/2 Heisenberg model with periodic boundary conditions on a 32 site chain at different \(k\)-points using LR-VMC. The results are shown in the left panel of Fig. \ref{fig:heisenberg_spec}. We used time-evolving block decimation to obtain the reference spectra. For LR-VMC, we used complex MLP states with an increasing number of hidden neurons with translational symmetry. Note that since this is a stoquastic Hamiltonian, it is possible to use a real MLP with the signs fixed by the Marshall sign rule. But we found the use of complex MLPs to provide faster convergence of the spectra with respect to the number of parameters. We show calculations with hidden neuron densities \(\alpha = 3\) and 5, demonstrating the convergence of the spectra with increasing number of parameters. We see good agreement between the LR-VMC and DMRG with some noticeable deviations especially at \(k=\pi/2\) and \(k=3\pi/4\). The large memory requirements of our current implementation limit the number of hidden neurons we can use in these calculations. With use of direct methods, we believe it will be possible to obtain converged results. We note that these results show an improvement over previously reported NQS calculations for this system in Ref. \citenum{hendry2021chebyshev}.   

The power of the methods developed in this work stems from the ability to treat the coupling of localized degrees of freedom to phonons. Thus we also studied the Heisenberg-bond model\cite{sandvik1997quantum} which includes a phonon modulated exchange interaction given as
\begin{equation}
   H = J\sum_{\langle ij\rangle} \left(1-g(b^{\dagger}_{ij}+b_{ij})\right)\mathbf{S}_i\cdot\mathbf{S}_{j} + \omega_0\sum_{ij}b^{\dagger}_{ij}b_{ij},
\end{equation}
where \(b_{ij}\) are phonon destruction operators residing on the lattice bonds. We again set \(J=1\). The dynamical spin structure factor for this model on a \(10\times 10\) lattice is shown in the right panel of Fig. \ref{fig:heisenberg_spec}. We used antiperiodic boundary conditions for this lattice. The results were obtained using LR-VMC with a complex MLP state with a density of hidden neurons \(\alpha=2\). We see a magnon band similar to the bare Heisenberg model. Fitting the linear spin wave theory result to the peaks in the LR-VMC magnon dispersion, we find \(J_{\text{eff}} = 1.47 J\), indicating a renormalized exchange interaction. We note that this scaling of \(J\) also partially reflects the underestimation of magnon energies by the linear spin wave theory. There is an additional structure around the optical phonon frequency \(\omega_0\) on top of the magnon spectrum due to the spin-phonon interaction. Thus for this set of parameters, the ground state is an antiferromagnet with renormalized exchange interactions due to the phonon coupling. We emphasize that the approach outlined here gives access to spectral properties of systems in two and higher dimensions coupled to phonons which are challenging for other numerical approaches like DMRG. We leave a detailed study of different phases of this system over the full parameter space to future work.


\begin{figure*}
   \centering
   \subfloat[Hubbard chain]{
   \includegraphics[width=0.9\columnwidth]{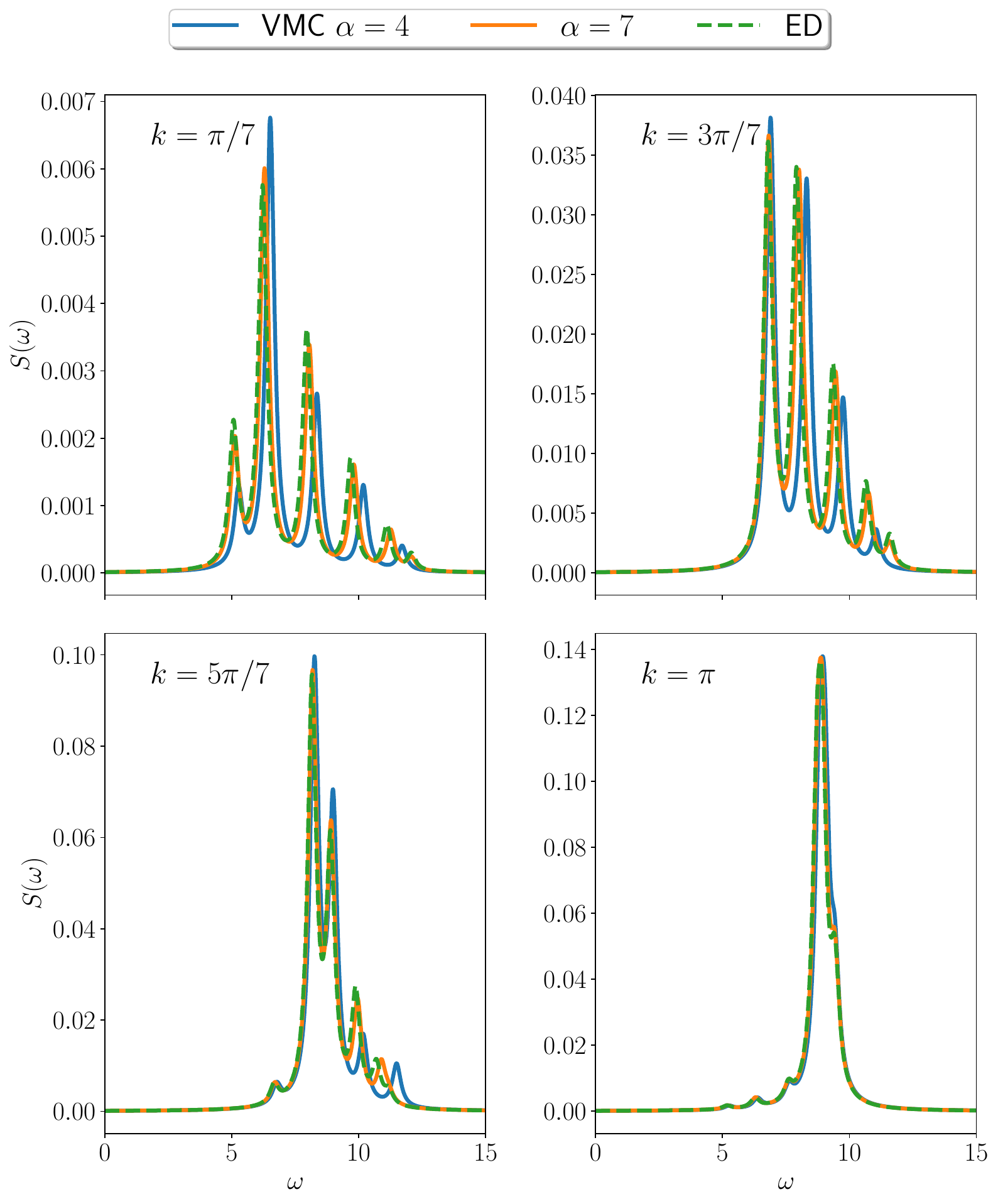}
   }\qquad
   \subfloat[Hubbard-Holstein chain]{
      \includegraphics[width=0.9\columnwidth]{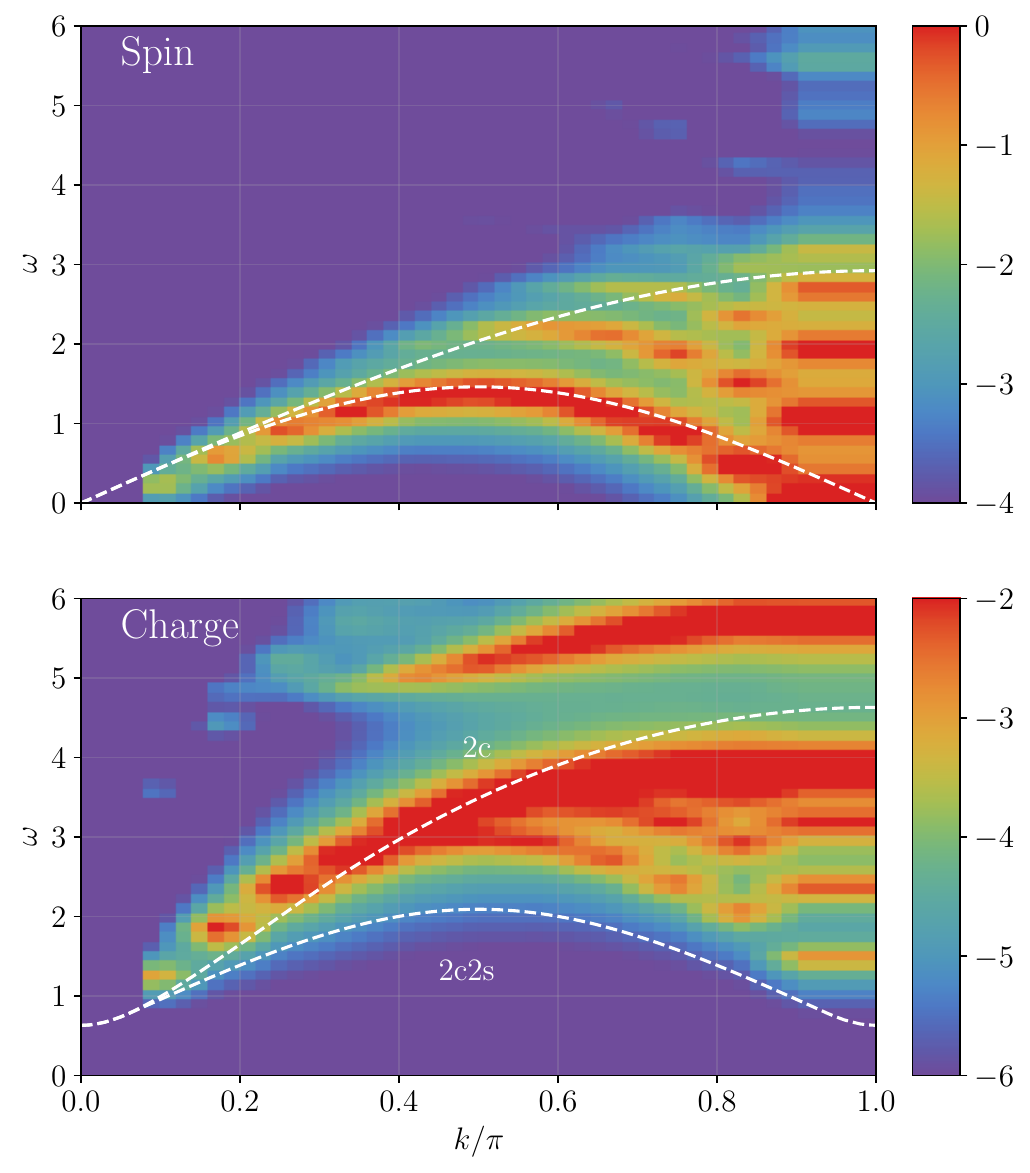}
   }
   \caption{(a) Convergence of the LR-VMC dynamical charge structure factor for the half-filled Hubbard model (\(U = 8\)) on a 14 site chain at different \(k\)-points with a broadening of \(\eta=0.2\). \(\alpha\) is the density of hidden neurons. (b) Spin and charge structure factors for the Hubbard-Holstein model for a half-filled 30 site chain with \(\omega_0=5, \lambda=0.25\) and \(U=4\). The trivial peak in the charge structure factor at \(k=\omega=0\) has been removed to enhance clarity. The dashed white lines show the Bethe ansatz\cite{essler2005one} bounds for the Hubbard model with \(U_{\text{eff}}=U-4\lambda=3\). For the spin structure factor, they mark the bounds of the two-spinon continuum. In the charge sector, they indicate the onset of the two-holon two-spinon (2c2s) and two-holon (2c) continuua.}\label{fig:hubbard_spec}
\end{figure*}


\subsubsection{Dynamical properties of the Hubbard-Holstein model}\label{sec:hubbard_holstein_spectra}
The Hubbard-Holstein model allows us to study the effect of eph coupling on charge dynamics in addition to the spin dynamics. Besides the dynamical spin structure factor, defined in the same way as for the Heisenberg model (Eq. \ref{eq:spin_structure_factor}), we consider the charge structure factor defined as
\begin{equation}
     S_{\rho}(k,\omega) = -\frac{1}{\pi}\text{Im}\expecth{\psi_0}{N_{-k}\frac{1}{\omega-H+E_0+i\eta}N_k}{\psi_0},
\end{equation}
where \(\ket{\psi_0}\) and \(E_0\) are the ground state wave function and energy, respectively. We first perform benchmark calculation on the purely electronic Hubbard model without eph coupling, where it is easier to obtain near-exact reference results for systems of nontrivial sizes. The left panel of Fig. \ref{fig:hubbard_spec} shows the convergence of the charge structure factor for the half-filled 14 site Hubbard model with \(U=8\) using LR-VMC. We compare our results to the reference ED values presented in Ref. \citenum{ido2020charge}. We confirmed that we obtain identical results for ED at \(k=\pi\) using our code. We used a \(S_z\) and momentum projected NQS-Jastrow GHF wave function. This represents a challenging regime for the LR-VMC method as it requires a good description of higher energy charge excitations. Despite this we see good agreement with the ED results for \(\alpha=7\) which uniformly improves upon the \(\alpha=4\) calculation.   

In the right panel of Fig. \ref{fig:hubbard_spec} we present dynamical spin and charge structure factors for the half-filled Hubbard-Holstein model on a 30 site chain. For the electronic part, we used the same wave function as the Hubbard model calculation, with phonons correlated in the NQS Jastrow factor. For the Hamiltonian parameters used here, the system is in the Mott insulating phase with quasi-antiferromagnetic ordering. In this phase, the spin spectrum is gapless and can be described by a two-spinon continuum, with a bulk of the spectral weight around \(k=\pi\). The charge spectrum, on the other hand, is gapped and consists of two-doublon excitations. There is also a weak contribution to the structure factor due to two-doublon two-spinon excitations.\cite{essler2005one} The spectra for the Hubbard-Holstein model shown in the figure follow this expected behavior and are in good agreement with those reported in Ref. \citenum{hohenadler2013excitation} obtained using the continuous time Monte Carlo method at a small but finite temperature. In the charge spectrum, we also see a feature at higher energies due to phonon excitations.

\section{Conclusion}\label{sec:conclusion}
In this paper, we have developed and examined the performance of neural quantum states for describing the effects of eph coupling in a wide class of models. Within these models, we considered different types of lattice models including diagonal and off-diagonal couplings. We considered different types of eph couplings, dimensionality, and the interplay of electron-electron interactions with eph coupling. In nearly all cases we found NQS to be able to describe ground state correlations accurately and efficiently. In extreme cases like the strong coupling regime of the SSH model, the NQS approach has some difficulty in describing the ground state correlations, but for polaron problems, it is possible to systematically obtain more accurate answers at the expense of a larger computational cost. We have also applied our methodology to calculate the hole polaron binding energy in LiF, demonstrating the possibility of using NQS to perform \textit{ab initio} calculations. Lastly, we studied a linear response strategy to calculate spectral properties based on NQS. This approach is attractive since it only requires a nonlinear stochastic optimization of the ground state, with the tangent space of the parameter manifold naturally serving as the response space. We showed that low-lying excitations can be well described in this framework without the need for manually constructing excited states. The ability to describe spectral properties accurately offers a sizeable advantage over imaginary time approaches which require analytic continuation for this task. 

Our work opens up many avenues of future research. Applications to more complex \textit{ab initio} systems can be enabled by exploiting the locality of interactions and low-rank properties in the Hamiltonian.\cite{luo2024data} Integrating semiclassical methods to account for acoustic phonons would allow the incorporation of these slow degrees of freedom more efficiently. Employing more sophisticated neural network architectures should facilitate the use of more efficient representations of the eph correlations. Enhancements in the implementation of dynamical calculations will enable the study of finite temperature transport and spectral properties in \textit{ab initio} systems. Lastly, leveraging strategies developed for describing electronic correlation in NQS, we also anticipate exploring the interplay between eph and electronic correlations in more realistic and complex models of strongly correlated electronic systems. Some of these directions will be explored in the immediate future. 

\section*{Acknowledgments}
We thank Marco Bernardi and Yao Luo for their helpful assistance in parametrizing the LiF \textit{ab initio} model. A.M. thanks Arkajit Mandal and Zhihao Cui for the useful discussions. A.M. and D.R.R. were partially supported by NSF CHE-2245592. P.J.R. acknowledges support from the National Science
Foundation Graduate Research Fellowship under Grant
No. DGE-2036197. This work used the Delta system at the National Center for Supercomputing Applications through allocation CHE230028 from the Advanced Cyberinfrastructure Coordination Ecosystem: Services and Support (ACCESS) program, which is supported by National Science Foundation grants \#2138259, \#2138286, \#2138307, \#2137603, and \#2138296.

\appendix
\section{Variance extrapolation of NQS energies}\label{app:variance_extrapolation}
In some cases, achieving convergence of energies with respect to the number of parameters in the NQS is challenging due to difficulties in the optimization of states with a large number of parameters. The most challenging cases of this class are models where the eph coupling strongly modulates the electron hopping. In these cases, we employ the technique of variance extrapolation to estimate the exact energy using a series of approximate calculations. This extrapolation is based on the rationale that since the exact ground state has zero energy variance (\(\langle H^2\rangle - \langle H\rangle^2\)), more accurate wave functions usually have lower energy variance in addition to lower variational energies. We use a linear fit to energies against variance to estimate the zero variance value.  

\begin{figure}
   \includegraphics[width=0.5\textwidth]{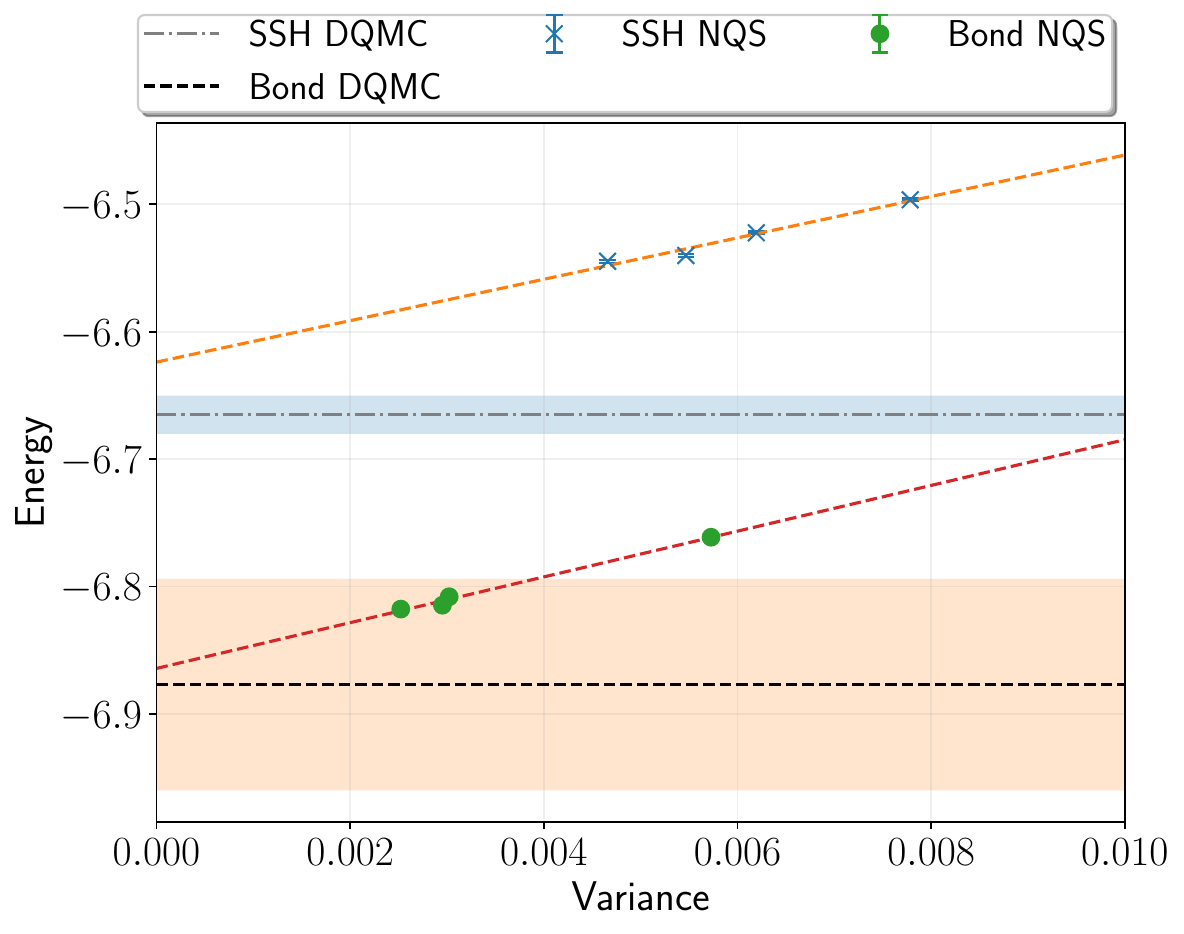}
   \caption{Variance extrapolation of NQS polaron energies for the  2D SSH (\(\omega_0=3, \lambda=1.82\)) and Bond (\(\omega_0=3, \lambda=1.45\)) models. States with number of hidden neurons given by \(n_H=50, 80, 120, 150\) were used to calculate the energies shown. Shaded regions show the stochastic errors on DQMC energies.}\label{fig:varext}
\end{figure}

Fig. \ref{fig:varext} shows the variance extrapolation of two-dimensional Bond and SSH model polaron ground state energies. DQMC energies taken from the work of \citet{zhang2021peierls} are shown for comparison. We chose values of coupling leading to similar energies in the two models to highlight the differences in convergence with respect to the number of hidden neurons. We see a slower convergence for the SSH model in the intermediate and strong coupling regime compared to the Bond model. The variance extrapolated NQS energy is within \(1\%\) of the binding energy obtained by DQMC. We note that DQMC has a sign problem for the SSH model that also renders convergence challenging for larger couplings.

\section{Sampling the Hamiltonian and overlap matrices}\label{app:lr_vmc_sampling}
Here we provide a more detailed description of the sampling of Hamiltonian and overlap matrices used in LR-VMC (see Eq. \ref{eq:lr_vmc}). One way to sample these matrix elements is by sampling the amplitude square of the ground state as

\begin{equation}
   \begin{split}
      \mathbf{H}_{\mu\nu} &= \frac{\expecth{\psi_{\mu}}{H}{\psi_{\nu}}}{\inner{\psi_0}{\psi_0}} = \sum_{w}\frac{|\inner{\psi_{0}}{w}|^2}{\inner{\psi_0}{\psi_0}}\frac{\inner{\psi_{\mu}}{w}}{\inner{\psi_0}{w}}\frac{\expecth{w}{H}{\psi_{\nu}}}{\inner{w}{\psi_0}},\\
      \mathbf{S}_{\mu\nu} &= \frac{\inner{\psi_{\mu}}{\psi_{\nu}}}{\inner{\psi_0}{\psi_0}} = \sum_{w}\frac{|\inner{\psi_{0}}{w}|^2}{\inner{\psi_0}{\psi_0}}\frac{\inner{\psi_{\mu}}{w}}{\inner{\psi_0}{w}}\frac{\inner{w}{\psi_{\nu}}}{\inner{w}{\psi_0}}.
   \end{split}\label{eq:lr_vmc_gs}
\end{equation}
This sampling method naturally follows from the ground state energy sampling approach, but it has the following issue. Because the configurations \(\ket{w}\) are drawn from the ground state distribution, they do not necessarily have substantial support on the tangent space states \(\ket{\psi_{\mu}}\). While this does not lead to the infinite variance problem seen in continuum simulations, for the discrete case one obtains high-variance estimates due to the ratios \(\frac{\inner{\psi_{\mu}}{w}}{\inner{\psi_0}{w}}\) becoming large for certain configurations, making the method statistically inefficient. 

One way to mitigate this problem is to use a different sampling function. This has been recognized in various QMC excited state studies \cite{ceperley1988calculation,filippi2009absorption,li2010variational}. Termed reweighting in Ref. \citenum{li2010variational}, this method uses the following sampling approach:

 \begin{equation}
   \begin{split}
      \mathbf{H}_{\mu\nu} &= \frac{\expecth{\psi_{\mu}}{H}{\psi_{\nu}}}{Z} = \sum_{w}\frac{Z_w}{Z}\frac{|\inner{\psi_0}{w}|^2}{Z_w}\frac{\inner{\psi_{\mu}}{w}}{\inner{\psi_0}{w}}\frac{\expecth{w}{H}{\psi_{\nu}}}{\inner{w}{\psi_{0}}},\\
      \mathbf{S}_{\mu\nu} &= \frac{\inner{\psi_{\mu}}{\psi_{\nu}}}{Z} = \sum_{w}\frac{Z_w}{Z}\frac{|\inner{\psi_0}{w}|^2}{Z_w}\frac{\inner{\psi_{\mu}}{w}}{\inner{\psi_0}{w}}\frac{\inner{w}{\psi_{\nu}}}{\inner{w}{\psi_{0}}},\\
      Z &= \sum_w Z_w = \sum_w \sum_{\mu} |\inner{\psi_{\mu}}{w}|^2.\\
   \end{split}\label{eq:lr_vmc_rew}
\end{equation}
Thus the configurations \(\ket{w}\) are sampled according to the distribution \(p(w) \propto \sum_{\mu} |\inner{\psi_{\mu}}{w}|^2\), which ensures sampling of configurations important for describing the excited states. The cost scaling of reweighted sampling is the same as the ground state sampling. 

\begin{figure}
   \centering
   \includegraphics[width=0.5\textwidth]{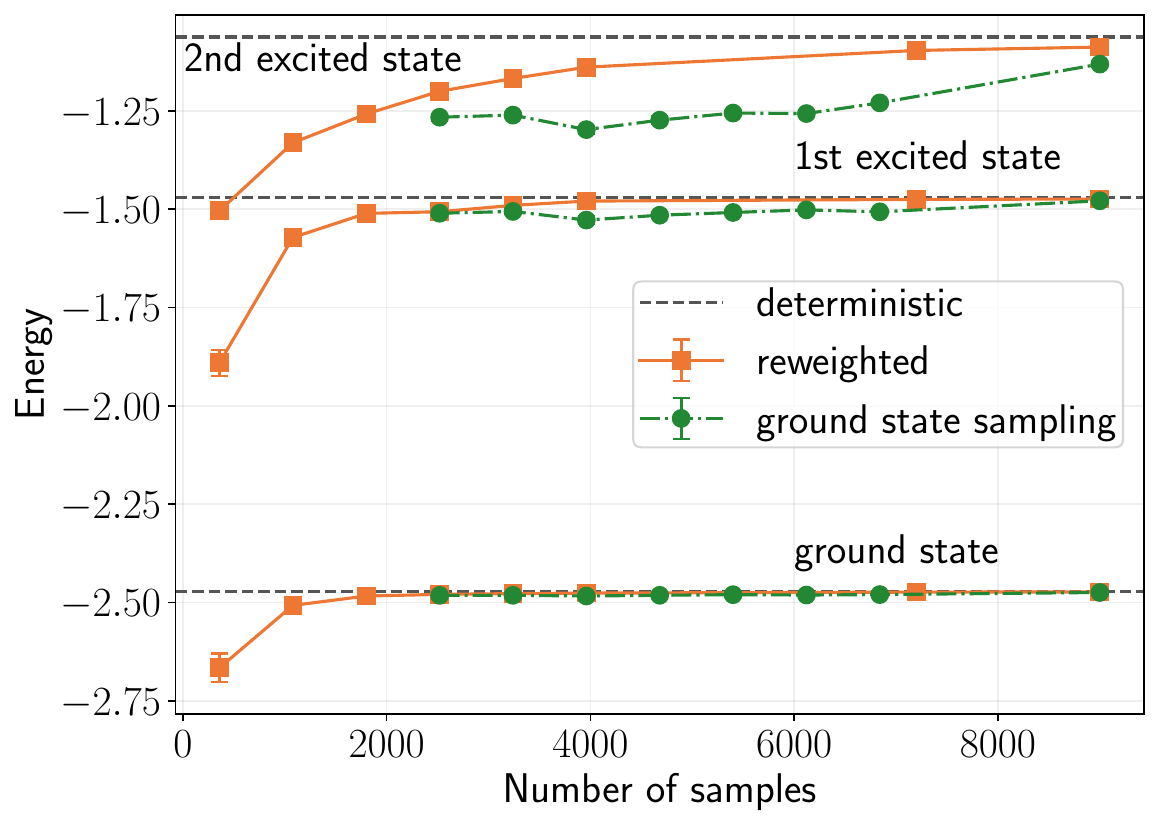}
   \caption{Energies of low-lying states of the Holstein polaron (\(\omega_0=1\), \(\lambda=1\), 6 sites, 5 maximum phonons) obtained using two LR-VMC sampling methods with different number of samples. The LR-VMC energies were obtained by averaging 100 independent calculations.}\label{fig:lr_vmc_sampling}
\end{figure}

The cost of constructing the \(\mathbf{H}\) and \(\mathbf{S}\) matrices scales as \(O(N_p^2N_s)\), where \(N_p\) is the number of parameters and \(N_s\) is the number of samples. This cost can be reduced by using a direct method, which only samples the action of these matrices onto vectors. This has been noted in several previous works mainly in the context of optimization methods\cite{neuscamman2012optimizing,sabzevari2020accelerated}. Consider the following expression for the action of the overlap matrix onto a vector \(\mathbf{x}\):

\begin{equation}
   \sum_{\nu}\mathbf{S}_{\mu\nu}x_{\nu} = \sum_{w}\frac{Z_w}{Z}\frac{|\inner{\psi_0}{w}|^2}{Z_w}\frac{\inner{\psi_{\mu}}{w}}{\inner{\psi_0}{w}}\left(\sum_{\nu}\frac{\inner{w}{\psi_{\nu}}}{\inner{w}{\psi_0}}x_{\nu}\right).
\end{equation}
The cost of this calculation scales as \(O(N_pN_s)\). The action of the Hamiltonian matrix can be similarly sampled. Iterative solvers can then be used to obtain spectral information of the system using only matrix vector products. In particular, the Chebyshev expansion-based kernel polynomial method\cite{weisse2006kernel} allows the calculation of various dynamical correlation functions including at finite temperatures.

The statistical performance of the two approaches is shown in Fig. \ref{fig:lr_vmc_sampling}. We performed these calculations on a small Holstein chain with a truncated phonon space to allow deterministic evaluation of the spectrum in LR-VMC, which serve as reference values. We restricted the LR-VMC sampling to the same truncated Hilbert space for the sake of this comparison. While Eqs. \ref{eq:lr_vmc_gs} and \ref{eq:lr_vmc_rew} yield unbiased estimates of the \(\mathbf{H}\) and \(\mathbf{S}\) matrix elements, their eigenvalues are biased, as they are nonlinear functions of the matrices.\cite{blunt2018nonlinear} Using each sampling approach, we calculated averages of 100 independent calculations using various numbers of samples to compute energies of the low-lying states. As expected, there is a bias in the obtained energies for a small number of samples, which decreases systematically as we increase the number of samples in the case of reweighted sampling. On the other hand, for ground state sampling we see a persistent large bias that is nearly unchanged from around 2000 to 6000 samples and suddenly decreases for 8000 samples. This is indicative of ergodicity issues in the ground state sampling approach to LR-VMC. Here, configurations with large contributions to the excited states do not get sampled often enough if the number of samples is small, leading to a large bias, especially in the excited state energies. We also find that numerical instabilities due to linear dependencies in the basis set \cite{blunt2018nonlinear, lee2021spectral} are greatly reduced due to reweighted sampling. 

\providecommand{\latin}[1]{#1}
\makeatletter
\providecommand{\doi}
  {\begingroup\let\do\@makeother\dospecials
  \catcode`\{=1 \catcode`\}=2 \doi@aux}
\providecommand{\doi@aux}[1]{\endgroup\texttt{#1}}
\makeatother
\providecommand*\mcitethebibliography{\thebibliography}
\csname @ifundefined\endcsname{endmcitethebibliography}
  {\let\endmcitethebibliography\endthebibliography}{}

\end{document}